\documentclass[sigconf]{acmart}

\usepackage{multirow}

\AtBeginDocument{%
  }

\copyrightyear{2026}
\acmYear{2026}
\setcopyright{cc}
\setcctype{by}
\acmConference[UIST '26]{The 39th Annual ACM Symposium on User Interface Software and Technology}{November 02--05, 2026}{Detroit, MI, USA}
\acmBooktitle{The 39th Annual ACM Symposium on User Interface Software and Technology (UIST '26), November 02--05, 2026, Detroit, MI, USA}
\acmDOI{10.1145/3830398.3830498}
\acmISBN{979-8-4007-2856-3/2026/11}

\begin{document}

\title{From Scaffolding to Internalization: Enhancing CPR Training with In-Situ Visualization and Kinesthetic Feedback}

\author{Jiahe Dong}
\authornote{Both authors contributed equally to this research.}
\email{dongjh2025@shanghaitech.edu.cn}
\orcid{0009-0003-1537-0717}
\affiliation{
\institution{School of Information Science and Technology, ShanghaiTech University}
  \city{Shanghai}
  \country{China}
}

\author{Shuhao Zhang}
\authornotemark[1]
\email{zhangshh12024@shanghaitech.edu.cn}
\orcid{0009-0008-1933-1869}
\affiliation{
\institution{School of Information Science and Technology, ShanghaiTech University}
  \city{Shanghai}
  \country{China}
}

\author{Yutao Ming}
\email{mingyt2025@shanghaitech.edu.cn}
\orcid{0009-0001-3220-7802}
\affiliation{
  \institution{School of Creativity and Art, ShanghaiTech University}
  \city{Shanghai}
  \country{China}
}

\author{Jinkai Zhang}
\email{zhangjk2023@shanghaitech.edu.cn}
\orcid{0009-0006-1097-1574}
\affiliation{
  \institution{School of Information Science and Technology, ShanghaiTech University}
  \city{Shanghai}
  \country{China}
}

\author{Yurui Zhang}
\email{yuruizhang105@gmail.com}
\orcid{0009-0006-1454-6030}
\affiliation{
    \institution{Institute of Microelectronics, University of Macau}
    \city{Macao}
    \country{China}
}

\author{Quan Li}
\authornote{Corresponding Author.}
\email{liquan@shanghaitech.edu.cn}
\orcid{0000-0003-2249-0728}
\affiliation{
\institution{School of Information Science and Technology, ShanghaiTech University}
  \city{Shanghai}
  \country{China}
}

\renewcommand{\shortauthors}{Dong and Zhang et al.}

 
\begin{abstract}
CPR training requires learners to not only understand explicit procedural targets, such as compression depth and rate, but also to internalize these targets as stable psychomotor skills. However, existing CPR training systems often rely on feedback presented outside the action space, which divides learners' attention between performing compressions and monitoring external guidance. This separation weakens the coupling between action and bodily sensation and may lead to an over-reliance on external feedback, compromising skill retention once support is removed. To address this challenge, we conducted a formative study with novice trainees and certified BLS instructors, from which we derived three design goals: embedding feedback within the task space, providing active kinesthetic guidance, and gradually fading assistance based on learning phases. Informed by these insights, we designed \textit{Kinesthetic-CPR}, a stage-adaptive multimodal mixed reality CPR training system, and evaluated it in a controlled user study across two sub-studies (N = 60). This work offers design implications for CPR training systems that aim to better support skill retention.
\end{abstract}

\ccsdesc[500]{Human-centered computing~Haptic devices}

\keywords{CPR Training, Kinesthetic Feedback, Mixed Reality, In-Situ Visualization, Motor Learning}

\begin{teaserfigure}
  \includegraphics[width=\textwidth]{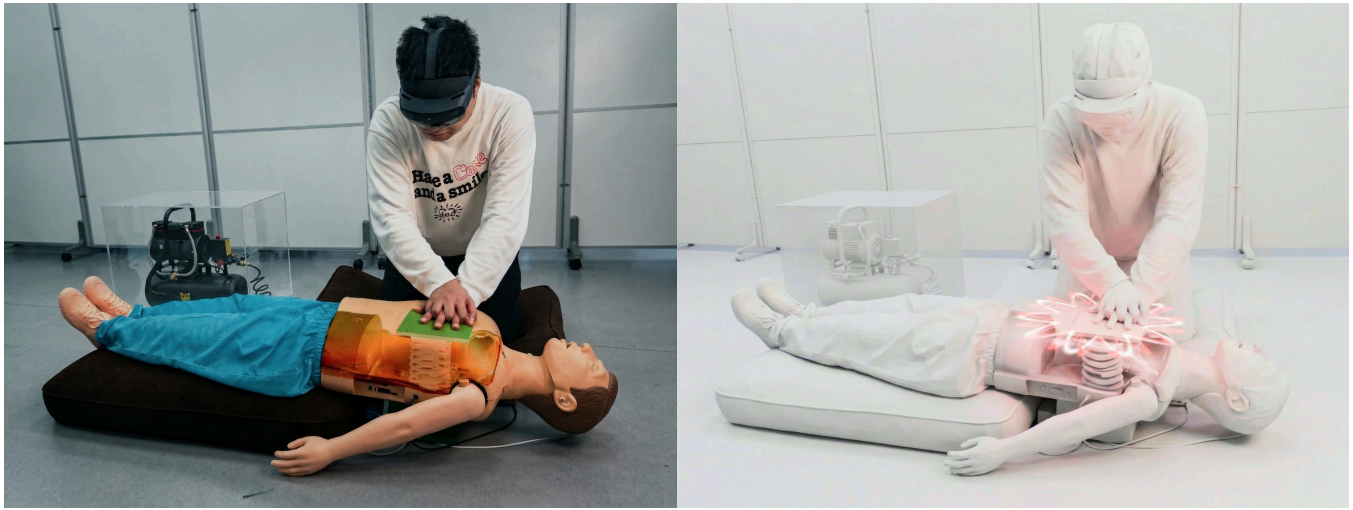}
  \caption{\textit{Kinesthetic-CPR}: a stage-adaptive multimodal mixed reality (MR) system that provides active physical guidance and spatially aligned visual feedback for CPR training. The system operates through two synchronized channels. (Left) Kinesthetic Channel: A custom pneumatic manikin regulates mechanical resistance to deliver physically grounded guidance. The air compressor (concealed during the user study to eliminate noise distractions) dynamically adjusts the internal cylinder's pressure. (Right) Visual Channel: An MR headset renders an in-situ visual feedback ring directly onto the compression area, aligning the digital guidance with the user's physical action space.}
  \Description{Side-by-side illustration of the Kinesthetic-CPR system during chest-compression training. On the left, a trainee wearing an MR headset performs CPR compressions on a supine manikin placed on a floor mat. The manikin’s chest area is shown with a translucent overlay revealing the internal pneumatic mechanism, and an external air compressor is positioned behind the manikin to indicate active mechanical control. On the right, the same training scene is rendered in a stylized mixed-reality view, where a glowing ring and radiating waveform are projected directly over the manikin’s chest to show spatially aligned in-situ visual feedback at the compression site. Together, the two panels depict the system’s synchronized kinesthetic guidance and visual feedback channels.}
  \label{fig:teaser}
\end{teaserfigure}

\maketitle

\section{Introduction}

\par Sudden cardiac arrest (SCA) remains a major global cause of mortality~\cite{marijon2023lancet}, making high-quality cardiopulmonary resuscitation (CPR) a critical determinant of survival~\cite{michelland2023association}. While clinical guidelines provide explicit performance criteria such as a compression depth of 5–6 cm and a rate of 100–120 compressions per minute~\cite{michelland2023association}, novices consistently struggle to achieve and maintain these metrics during hands-on practice~\cite{yeung2009use,yang2012systematic}. The underlying difficulty is that effective CPR is not merely a declarative task, but a complex psychomotor skill requiring precise regulation of force, rhythm, and body posture~\cite{sense2016interactions,perkins2015european}. For novice trainees, the primary bottleneck is not memorizing numerical targets, but the absence of internalized proprioceptive reference points to guide their physical performance~\cite{burkhardt2014effect}. Consequently, training systems must actively support learners in bridging the gap between explicit cognitive knowledge and implicit motor control to construct a robust kinesthetic schema.

\par To support this transformation, CPR training methodologies have evolved toward sensor-equipped manikins that provide real-time performance feedback via external displays or auditory cues~\cite{schmidt1975schema,baldi2017real,white2017measuring,gruenerbl2018training,di2020real,himane2025stayin,liu2025electrotactile}. While this feedback effectively improves compression quality during active use, a critical objective of training is to ensure that learners retain the skill and perform it independently once the feedback is removed. Despite their immediate performance benefits, however, these systems present three fundamental limitations that hinder motor internalization. First, feedback is typically presented outside the task space, forcing learners to constantly shift attention between their hands and external interfaces—a spatial separation that increases cognitive load (\textbf{Gap1})~\cite{wagner2022visual,chandler1992split}. Second, although recent extended reality (XR) systems attempt to mitigate this by superimposing visual cues directly onto the manikin~\cite{park2013projected}, this in-situ feedback remains inherently symbolic. Because real CPR relies heavily on the mechanical resistance of the chest, symbolic visual cues fail to couple feedback with learners' proprioception (\textbf{Gap2})~\cite{schmidt2008motor,adams1976issues}. Active kinesthetic guidance is therefore required to help novices form reliable perceptual references. Finally, existing systems predominantly rely on static feedback intensities. Continuous high-intensity scaffolding can foster feedback dependency, undermining autonomous performance and long-term skill retention once assistance is removed, revealing a critical lack of stage-adaptive fading strategies (\textbf{Gap3})~\cite{schmidt1989summary,fitts1967human,lim2021rapid}.

\par Based on these three identified gaps, we argue that the core problem of current CPR training systems is not the insufficiency of individual feedback modalities, but rather the absence of a cohesive design perspective that systematically addresses three interrelated dimensions: the spatial alignment between feedback and the user's action space, the coupling of feedback with proprioceptive cues, and the temporal modulation of feedback intensity across learning stages. In other words, to facilitate the transition from explicit knowledge to implicit motor control, training systems must be reconceptualized as a holistic design space that supports the internalization of psychomotor skills. This motivates our first research question: \textbf{RQ1:} \textit{What are the key design requirements and constraints for CPR training systems aimed at fostering psychomotor skill internalization within this design space?}

\par To answer \textbf{RQ1}, we conducted a formative study with certified BLS instructors and novice trainees to examine challenges in CPR feedback, action correction, and skill retention. The study revealed three design goals for supporting motor internalization: aligning feedback with the action space, establishing action–sensation mappings, and fading support as competence develops. These findings led to \textbf{RQ2:} \textit{How can in-situ visual feedback, active kinesthetic guidance, and stage-adaptive feedback scheduling be integrated into a unified CPR training system to support motor internalization?}

\par To answer \textbf{RQ2}, we designed \textit{Kinesthetic-CPR}, an MR-based training system that combines in-situ visual feedback, active kinesthetic guidance, and stage-adaptive fading. The system projects compression cues onto the manikin’s chest, provides mechanical guidance for depth and rhythm, and progressively reduces support across training rounds.

\par To evaluate the system's effectiveness, we pose \textbf{RQ3}: \textit{To what extent can spatially aligned visual feedback, active kinesthetic guidance, and adaptive feedback fading improve CPR learning outcomes, including immediate performance, cognitive load, and skill retention?} To answer it, we conducted two interconnected sub-studies with 60 participants. The contributions of this study are threefold:

\begin{itemize}
    \item We conducted a formative study with certified BLS instructors and novice trainees, systematically identifying key design requirements and core challenges for supporting psychomotor skill internalization in CPR training.
    \item We designed an active kinesthetic intervention for CPR training, implemented through a custom pneumatic manikin that transforms chest resistance into a real-time corrective interface for both compression depth and rhythm, and integrated this mechanism with in-situ MR feedback and a predefined three-stage fading schedule within a unified training architecture.
    \item Through a controlled user study across two sub-studies ($N=60$), we evaluated how spatially aligned visual feedback, active kinesthetic guidance, and adaptive feedback fading influence immediate training performance, cognitive load, and skill retention.
\end{itemize}

\section{Related Work}

\subsection{Evolution of CPR Training Systems}
\par According to a consensus statement from the American Heart Association (AHA), key indicators of CPR quality—including compression depth, rate, and chest recoil—are insufficiently assessed by visual observation alone~\cite{meaney2013cardiopulmonary}. This limitation has led to the widespread adoption of instrumented manikins as the cornerstone of modern CPR training. These systems typically employ accelerometers, pressure sensors, or displacement sensors to capture kinematic and dynamic data during compressions, delivering knowledge of results to learners in real time. The evolution of CPR training technology has been largely defined by advancements in how such performance data is communicated. Early systems predominantly used auditory cues, such as metronomes, to guide rhythm~\cite{liu2025electrotactile}. Subsequently, visual displays and graphical interfaces became common, providing feedback on compression waveforms, depth, and rate~\cite{baldi2017real}. More recent research has explored non-visual modalities such as haptic vibrations and thermal signals to indicate compression deviations during practice~\cite{liu2025electrotactile,durai2019effect}. Meanwhile, Virtual Reality (VR) and Augmented Reality (AR) have been integrated into CPR training to enhance contextual fidelity, embed performance metrics, and support immersive practical exercises~\cite{semeraro2019back,sigrist2013augmented, leary2020pilot}.

\par Parallel to innovations in feedback delivery, the mechanical architecture of CPR manikins has advanced to better replicate the biomechanical properties of the human thorax under compression. Early manikins commonly utilized linear components such as single springs, which offered basic resistance but failed to mimic the nonlinear viscoelastic behavior of the chest during compressions~\cite{gruben1993sternal}. To enhance haptic realism, subsequent designs incorporated hybrid spring-pneumatic systems to approximate thoracic damping more accurately~\cite{nysaether2008manikins}. Others introduced adjustable stiffness or variable damping mechanisms to simulate differences in chest resistance due to age, body habitus, or pathological conditions~\cite{xie2009simulator,stanley2012recreating}. More recent high-fidelity manikins have integrated synthetic ribs, soft tissues, and organ analogs to provide tactile feedback that more closely resembles actual human anatomy~\cite {thielen2017innovative}.

\subsection{In-situ Feedback and Task-Space Alignment in XR-based Psychomotor Training}
\par In conventional CPR training, key feedback such as compression depth and rate is often presented on external displays or tablets, requiring learners to shift attention between the manikin, their hands, and the interface~\cite{baldi2017real}. This separation between information space and task space increases split attention and can impair real-time motor performance~\cite{chandler1992split}.

\par XR offers a promising approach to alleviate this spatial fragmentation~\cite{buchner2022impact}. By leveraging AR or MR, recent studies in both CPR and surgical training demonstrate that contextually anchored visualizations can significantly reduce gaze shifts and strengthen the perceptual link between feedback and the physical body~\cite{johnson2012eye}. Nevertheless, the implementation of XR in CPR training varies in its adherence to spatial and cognitive design principles. Many current systems simply transpose traditional graphical interfaces into floating virtual panels positioned near the manikin~\cite{ohshima2023mr}. While this reduces physical head movement, it fails to achieve true spatial alignment, as learners must still mentally map offset information onto their hands. Recent AR and MR motor-training systems have provided situated visual feedback, 3D avatar instructions, and automated coaching for sports and exercise~\cite{iannucci2023arrow,ihara2025video2mr, lee2026vistar}.

\par Beyond spatial considerations, the representational format of positional feedback also plays a critical role in high-stress training scenarios. Research indicates that in simulated emergencies, even minor cognitive delays can compromise performance, underscoring the need for rapidly interpretable visualizations~\cite{zhang2026seconds}. However, some XR training interfaces repurpose instructor-oriented dashboards for trainees, encoding depth and rate as analytical clinical waveforms or dense numerical arrays~\cite{tanaka2019effect}. Such representations increase cognitive load and lack the immediacy required for pre-attentive processing. Furthermore, continuously displayed visual overlays may obstruct the trainee's view of their hands and the patient's chest. To mitigate this, emerging XR designs emphasize adaptive rendering—for example, providing feedback only when performance deviates from a target range, thereby preserving visual clarity and leveraging negative space~\cite{zhang2026seconds}. Additionally, effective systems distinguish between procedural guidance and motor guidance, routing the former to peripheral head-up displays to avoid cluttering the central visual field~\cite{johnson2018holocpr}. While kinematic knowledge of performance benefits from strict spatial alignment with the chest, instructional prompts are better suited for non-obstructive peripheral presentation.

\subsection{Kinesthetic Feedback and Adaptive Fading Assistance in Psychomotor Learning}
\par Haptic feedback generally includes two complementary forms: tactile feedback, which acts on the skin surface, and kinesthetic feedback, which acts on muscles and joints through force~\cite{hamam2013effect}. In CPR training, prior work has primarily relied on tactile haptic cues such as vibration~\cite{sarac2026multimodal}. While such feedback can indicate whether compressions are too fast or too shallow, it does not directly guide the learner’s movement or help establish a proprioceptive reference for correct action. Beyond optimizing the spatial presentation of feedback, a complementary line of inquiry in psychomotor learning examines the interplay between feedback and proprioception. Motor learning theory suggests that skill acquisition involves gradually internalizing stable action–sensation mappings rather than merely understanding external performance outcomes~\cite{schmidt1975schema,adams1976issues}. In CPR, this means developing an internal sense of appropriate depth and rhythm through bodily interaction with chest resistance~\cite{odegaard2007chest}. Accordingly, an increasing number of studies have begun to focus on the role of active proprioceptive guidance in supporting psychomotor skill development~\cite{aman2015effectiveness}.

\par In haptic and force feedback research, active kinesthetic feedback has been applied to various skill-learning tasks, including rhythmic movement training~\cite{grindlay2008haptic}, handwriting and trajectory following~\cite{gutierrez2024immersive}, robot-assisted surgery, and motor rehabilitation~\cite{gomez2025simulation}. These studies indicate that when a system is capable of directly influencing limb movements through variable resistance, guiding forces, or haptic rhythms, learners more readily internalize the temporal and force patterns of target actions~\cite{sigrist2013augmented}.

\par However, such research also points out that the effectiveness of active guidance is not merely a matter of ``more is better''. The guidance hypothesis in motor learning suggests that while high-intensity external assistance improves immediate performance, its sustained presence can foster over-reliance on system-provided scaffolding. This, in turn, may undermine learners' capacity for independent performance once the assistance is withdrawn~\cite{schmidt1989summary}. In response, researchers in rehabilitation, haptic instruction, and robot-assisted learning have increasingly turned to adaptive strategies such as faded assistance and assist-as-needed paradigms~\cite{huegel2010progressive, anderson2013youmove}.

\section{Formative Study}
\par Despite prior work in CPR training technologies, in-situ XR feedback, and motor learning having identified limitations in spatial alignment, sensory coupling, and stage-adaptive support, there is limited empirical evidence from real instructional practice on how these limitations unfold in real settings~\cite{schipper2025technological}, how learners interpret and adjust their actions based on feedback, and how instructors perceive the development of haptic feel and the fading of learning scaffolds. To answer \textbf{RQ1}, we conducted a formative study combining observations of novice practice with interviews of CPR instructors. This approach aimed to identify key barriers in current instructional practice and derive design goals for the system.

\subsection{Setup}

\par A total of 14 participants took part in this formative study, including 6 certified BLS instructors (I1–I6) and 8 novice learners with no recent CPR training experience (T1–T8). Recruitment was conducted through posters distributed in collaborating hospitals and via snowball sampling. The study was approved by our Institutional Review Board, and all participants provided informed consent. All participants received approximately \$20 as compensation for their time. Detailed demographics are summarized in the Appendix (\autoref{tab:appendix_participant-info-formative}).

\par The study combined semi-structured interviews with observational sessions. During the observation session, novice learners were instructed to perform two minutes of continuous chest compressions on a training manikin equipped with a standard external visual feedback display. Feedback remained active for the first minute and was deactivated for the second. Observations focused on gaze patterns, movement correction strategies, and performance stability across conditions, followed by a 15-minute interview exploring their feedback reliance and kinesthetic experiences.

\par For the instructor group, we conducted 45--60-minute interviews exploring instructional challenges, physical correction strategies, learners' attentional patterns, and the tension between skill retention and feedback withdrawal. All sessions were audio-recorded and transcribed verbatim. Data were analyzed using thematic analysis~\cite{braun2006using}. Two researchers independently conducted open coding on the interview transcripts and observation notes, identifying recurring patterns related to attentional distraction, difficulties in movement correction, and dependence on external scaffolds. Through iterative comparison and discussion, these codes were progressively synthesized into the final qualitative findings.

\subsection{Findings from the Formative Study}

\par Through the thematic analysis of observational data and interview transcripts, we identified three primary limitations inherent in current CPR training paradigms. The following sections detail the specific barriers preventing novice learners from internalizing motor skills, directly informing our subsequent design goals.

\par \textbf{F1. External visual feedback exacerbated the spatial separation between task space and information space, hindering the development of situated perception.} External visual interfaces required learners to constantly shift attention between the physical task space and the display, causing discomfort upon removal. T4 noted, ``\textit{I wanted to know whether the force of my compressions was appropriate, so I kept sneaking glances at the screen.}'' T6 similarly remarked, ``\textit{Without the numerical indicators on the screen, I felt nervous because I could not tell whether my actions were accurate.}''  Instructors observed that this fragmented the task and weakened learners' awareness of the simulated emergency context. As I2 explained,  ``\textit{The learners' attention is entirely on the monitor, and they ignore the simulated patient under their hands. This kind of separation does not exist—and would not be acceptable—in a real emergency.}'' I4 similarly observed, ``\textit{You can clearly see that the more frequently learners shift their gaze, the worse their compression performance tends to be, because it becomes harder for them to maintain a continuous rhythm.}'' This spatial separation increased cognitive load and disrupted the perceptual coupling between visual cues and physical action.

\par \textbf{F2. Symbolic feedback lacked an embodied referent, limiting its translation into stable and immediate motor correction.} Symbolic feedback requires learners to cognitively decode information before inferring the appropriate muscular adjustments. T1 described this challenge: ``\textit{When I was told that my compression posture was incorrect, I could only try changing my force experimentally. It was very hard to immediately find the right posture that was both compliant and effort-efficient.}'' Instructors' accounts similarly highlighted the difficulty of teaching proper compression posture and force generation. I3 stated, ``\textit{Numbers cannot tell learners how their shoulders should drop or how their core should engage. When they lack internal muscle memory, warning signals only make them more flustered.}'' I4 added, ``\textit{In those situations, I usually have to correct them by physically guiding the movement, hoping they can actually feel what a correct compression is like.}'' These observations suggest that symbolic error indicators cannot provide novices with a proprioceptive reference for the target action. As a result, movement correction remains a trial-and-error process, limiting the development of a stable kinesthetic schema.

\par \textbf{F3. Static external scaffolds supported immediate performance but failed to facilitate motor skill internalization.} Novice learners consistently expressed strong uncertainty once the external screen was removed. T8 commented, ``\textit{After the screen was turned off, I could keep going based on short-term memory, but that short-term memory is fragile. Once I get interrupted, I cannot recover the rhythm.}'' Instructors indicated that static external cues sustain training performance but fail to prepare learners for independent execution. I5 observed, ``\textit{As long as there is a metronome, novices usually keep a very standard rhythm. But once that support is removed, their rate quickly drifts.}'' I1 elaborated, ``\textit{Depth is an even bigger problem. Once the support is removed, learners gradually fall back to their comfort zone in terms of compression depth.}'' Existing feedback mechanisms cause learners to develop responses to external cues rather than an internal proprioceptive memory, impeding true skill acquisition and autonomous proficiency.

\subsection{Design Goals}
\par Based on insights from the formative study, we derived three core design goals to guide the development of a system that supports the internalization of CPR psychomotor skills.

\par \textbf{DG1. Ground feedback directly within the physical action space.} To reduce cognitive load and mitigate the contextual detachment introduced by cross-space information integration, the system should deliver performance feedback directly at the operative region of the training manikin through in-situ visualization. This approach enables learners to access real-time information about compression depth, rate, and other performance metrics while maintaining continuous visual attention on the physical entity being acted upon, thereby supporting situated action and preserving the perceptual coupling between feedback and task (\textbf{F1}).

\par \textbf{DG2. Provide physically grounded, active kinesthetic guidance.} Since symbolic feedback alone is insufficient to support immediate and accurate motor correction, the system should provide mechanical intervention that acts directly on the learner's body. This kinesthetic guidance should move beyond merely signaling error states; it should enable learners to physically experience the target compression force and rhythm. In doing so, it helps establish a proprioceptive reference aligned with the desired action and facilitates the stabilization of action–sensation mapping (\textbf{F2}).

\par \textbf{DG3. Implement adaptive fading external scaffolding.} To address the limitation that static assistance can hinder skill internalization, the system should incorporate a dynamic feedback scheduling strategy. The intensity of external support should progressively decrease in response to the learner's real-time performance and stage of learning, enabling a smooth transition from strong external dependence to autonomous control grounded in an internal kinesthetic schema. This design supports long-term skill retention even after the system is removed (\textbf{F3}).

\section{Kinesthetic-CPR}
\par To operationalize the design goals (\textbf{DG1}--\textbf{DG3}) identified in the formative study, we designed and developed \textit{Kinesthetic-CPR}, a system that provides active physical guidance and spatially aligned visual feedback for CPR training. This approach is designed to support the internalization of psychomotor skills in high-pressure medical contexts. In this section, we describe the overall system architecture, the in-situ visual guidance interface, the active kinesthetic feedback mechanism, and the stage-adaptive fading strategy for skill retention.

\subsection{System Overview}
\par
\begin{figure}[h]
    \centering
    \includegraphics[width=\linewidth]{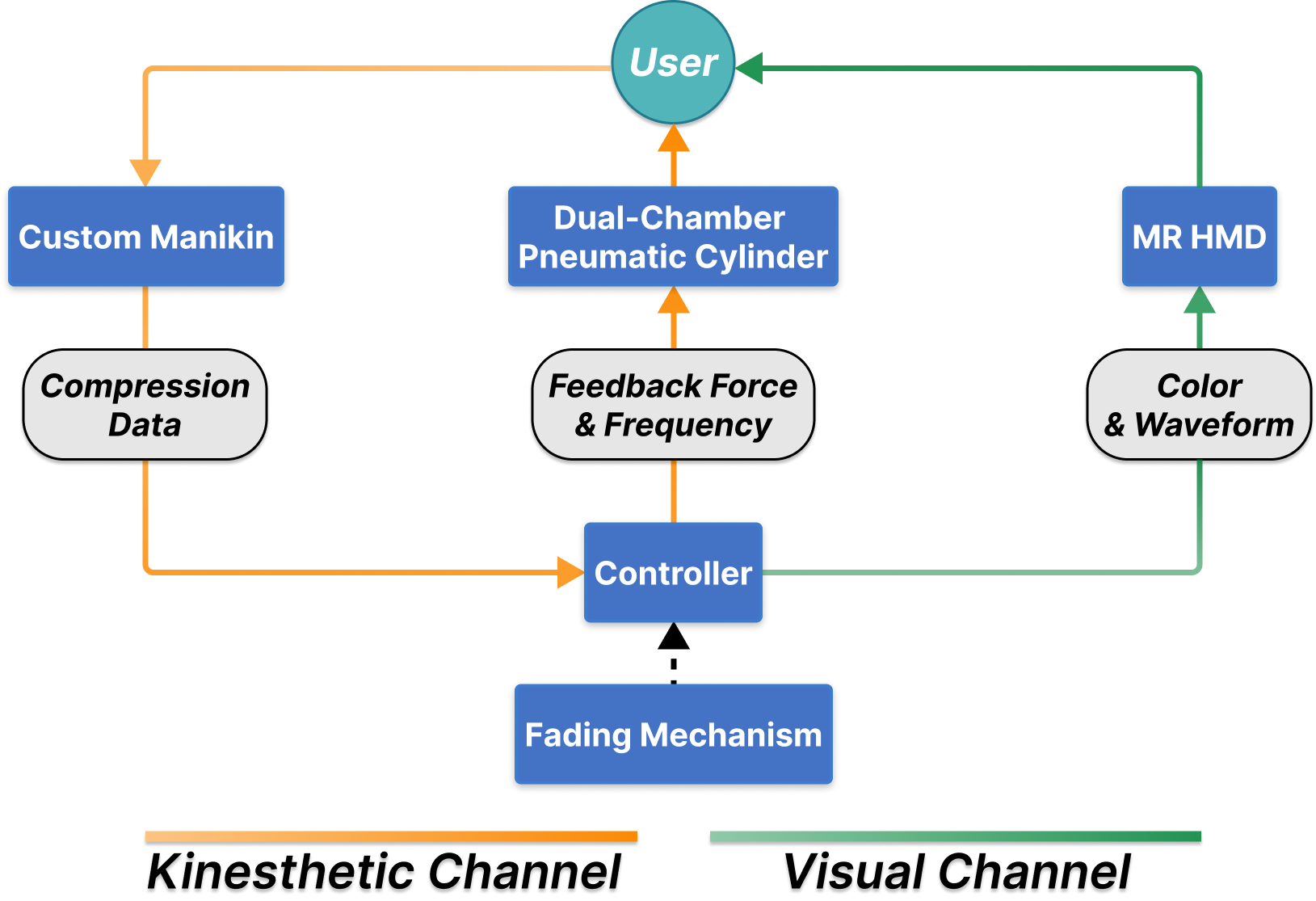}
    \caption{The workflow of \textit{Kinesthetic-CPR}. A central controller processes real-time compression data from the custom manikin to drive two feedback loops. The kinesthetic channel uses a dual-chamber pneumatic cylinder to adjust mechanical resistance and deliver force feedback. The visual channel uses an MR headset to display color and waveform metrics. Both modalities are modulated by a stage-adaptive fading mechanism that dynamically adjusts intervention intensity.}
    \Description{Block diagram of the Kinesthetic-CPR workflow organized into two parallel feedback channels. At the center, a controller receives compression data from a custom manikin and sends commands to both a dual-chamber pneumatic cylinder and an MR headset. In the kinesthetic channel, the pneumatic cylinder returns feedback force and frequency to the user through the physical manikin. In the visual channel, the MR headset presents color and waveform feedback to the user. A fading mechanism positioned below the controller modulates both channels, indicating stage-adaptive adjustment of feedback intensity over time. Orange arrows denote the kinesthetic channel, and green arrows denote the visual channel.}
    \label{fig:SystemWorkFlow}
\end{figure}

 The architecture of \textit{Kinesthetic-CPR} consists of three core modules: an MR visual front end, a custom pneumatic training manikin, and a closed-loop control backend (\autoref{fig:SystemWorkFlow}). During operation, a high-precision laser displacement sensor embedded beneath the manikin's chest continuously samples chest displacement at 100~Hz, from which compression depth and rate are derived (\autoref{fig:structure}). The resulting kinematic data stream is transmitted synchronously via an ESP32 microcontroller to the computational backend, where feature extraction and state assessment are performed.

\par Based on the incoming compression performance and the current training phase, the backend dynamically schedules feedback interventions across two modalities. In the visual channel, performance indicators are rendered directly onto the manikin's chest through a HoloLens 2~\cite{microsoft_hololens2} headset, providing spatially aligned feedback (\textbf{DG1}). In the physical channel, the microcontroller drives a proportional electro-pneumatic valve~\cite{smc_itv3050} to regulate the pressure inside the pneumatic cylinder chamber, thereby modulating the mechanical resistance of the manikin chest and delivering real-time rhythmic feedback (\textbf{DG2}). To prevent learner dependency on external assistance and to foster proprioceptive competence, both the visual and kinesthetic feedback modules are governed by a global stage-adaptive fading algorithm, which progressively reduces intervention intensity across successive training rounds (\textbf{DG3}).

\subsection{In-Situ Visual Guidance}
\par To address \textbf{DG1}, we designed an MR visual guidance interface for CPR chest compressions that embeds performance feedback directly into the task space where the user interacts with the manikin. Through \textit{in-situ visualization}, the interface spatially co-locates feedback with the compression site, reducing the need for attentional shifts between physical action and interface interpretation and preserving continuous \textit{situated perception}. This design directly responds to the issue of attentional separation identified in the formative study (\textbf{F1}).

\par The visual encoding avoids dense numerical panels or standalone dashboards, instead employing a compact visual language centered on a circular dynamic waveform to support rapid assessment of compression states. Following the design principles of situated analytics and embedded data representations in extended reality environments~\cite{willett2016embedded}, the system conformally maps two-dimensional visual elements directly onto the surface of the manikin's chest. Compared to three-dimensional floating visuals, this dimensionality-reduction and surface-mapping strategy effectively mitigates perspective distortion and visual occlusion, ensuring consistent spatial perception~\cite{marriott2018immersive,munzner2025visualization}. Technically, we implemented in-situ alignment using MRTK's Solver, which spatially couples the feedback interface to the learner's active compression hand. Because the hand remains directly above the compression site during CPR, the interface tracks the ongoing compression motion, thereby maintaining feedback near the action site and mitigating overlay jitter and spatial drift caused by rapid compressions and head movements.

\par
\begin{figure}[h]
    \centering
    \includegraphics[width=\linewidth]{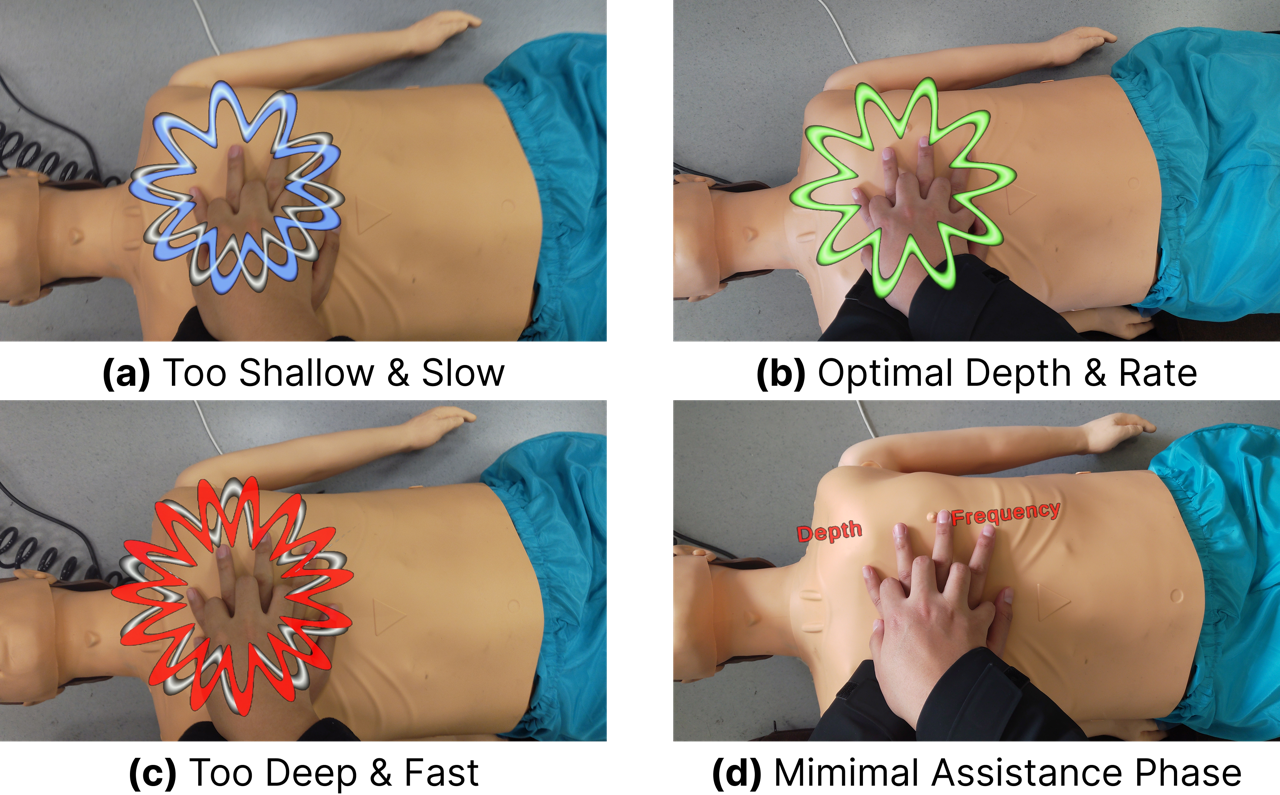}
    \caption{The in-situ MR visual guidance interface under different compression states. (a) Too shallow and slow: the ring turns blue, and the dynamic waveform is sparser than the reference. (b) Optimal performance: the ring turns green with the dynamic waveform perfectly aligned with the reference. (c) Too deep and fast: the ring turns red, and the waveform is denser than the reference. (d) Minimal assistance phase: continuous graphical elements are removed, leaving only textual warnings that flash upon detecting deviations.}
    \Description{Four-panel top-down view of the in-situ MR guidance interface projected onto a CPR manikin’s chest during compressions. Panel (a) shows a blue ring with a relatively sparse waveform, representing compressions that are too shallow and too slow. Panel (b) shows a green ring with a balanced waveform aligned to the target pattern, representing optimal depth and rate. Panel (c) shows a red ring with a denser waveform, representing compressions that are too deep and too fast. Panel (d) shows the minimal-assistance condition, in which the continuous ring visualization is removed and only small textual warnings appear near the hands when depth or frequency deviates from the target range.}
    \label{fig:MR_interface}
\end{figure}

 Given the inherently high cognitive load of cardiopulmonary resuscitation tasks, the interface leverages pre-attentive visual processing mechanisms~\cite{ware2019information,lu2020glanceable} to enable glanceable information retrieval. During prototyping, we explored encoding compression rate via texture flow speed, ring rotation, and visual vibration. We ultimately selected ring color and waveform density, as they represent depth and rate through visually separable channels, facilitate rapid comparison against categorical target ranges, and avoid introducing directional motion or transient visual disturbance. Variations in color and spatial density convey two core kinematic dimensions: compression depth and rate simultaneously. Compression depth is represented in real time by the color of the ring, allowing learners to quickly detect deviations while maintaining motor continuity. Compression rate is encoded as a waveform texture embedded within the ring's contour, where the real-time spatial frequency represents the current compression rhythm. To facilitate immediate comparison, a semi-transparent reference waveform overlaid on the ring indicates the target rhythm. When the peak distribution and visual density of the real-time waveform align with the reference, the compression rate is within the target range; a waveform that is too dense indicates an excessively fast rate, while a sparse waveform indicates a slow rate. \autoref{fig:MR_interface} illustrates the interface performance under different compression states. The interface is implemented in Unity and maintains a real-time connection with the backend via Holographic Remoting to support live rendering and state updates during training.

\subsection{Active Kinesthetic Feedback}
\par Although visual cues address spatial fragmentation, symbolic feedback still relies on cognitive interpretation. To promote the development of embodied kinesthetic schemata (\textbf{DG2}), we constructed a custom CPR manikin based on a hybrid spring-pneumatic architecture that provides physically grounded motion guidance while preserving biomechanical realism. The chest is built around a set of custom-calibrated mechanical springs, whose force-displacement profile was matched to a predefined anatomical reference to simulate the passive compliance of the thoracic cage. To reproduce the viscous damping characteristics of biological tissue, a dual-chamber pneumatic cylinder was integrated coaxially at the center of the spring assembly. The upper chamber connects to a proportional valve, providing natural air damping that varies with compression velocity. The lower chamber serves as the actuation unit for active haptic interaction and is directly connected to an electronically controlled proportional valve driven by an ESP32. By dynamically regulating the pressure in the lower chamber, the system superimposes variable active forces onto the baseline resistance, enabling physical intervention on both compression rhythm and depth.

\par
\begin{figure}[h]
    \centering
    \includegraphics[width=\linewidth]{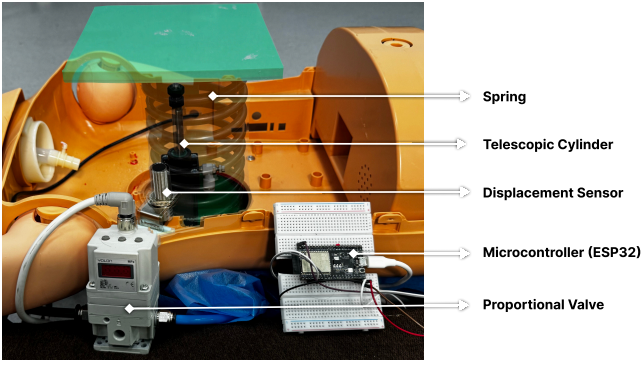}
    \caption{Hardware components of the custom training manikin. An ESP32 microcontroller processes real-time kinematic data from the displacement sensor. Based on this data, it controls a proportional valve to dynamically regulate the pressure within a coaxial telescopic cylinder, superimposing active kinesthetic guidance onto the baseline resistance provided by the mechanical springs.}
    \Description{Annotated photograph of the custom CPR training manikin with the outer shell opened to reveal its internal hardware. Labels identify a spring assembly, a telescopic cylinder, a displacement sensor, an ESP32 microcontroller mounted on a breadboard, and a proportional valve connected by tubing. The pneumatic and sensing components are integrated inside the manikin torso, illustrating how compression motion is measured and how air pressure is regulated to modify chest resistance during training.}
    \label{fig:structure}
\end{figure}

 Instead of implementing a high-fidelity dynamic controller, our system uses a lightweight closed-loop heuristic that transforms recent performance deviations into perceivable rhythmic pulses and resistance reshaping. To support closed-loop control, a laser displacement sensor mounted directly beneath the chest plate samples depth at 100~Hz. The resulting signal is used to estimate compression timing and peak displacement, allowing the system to provide motion-perceptual guidance separately for compression rate and depth. The detailed internal structural diagram is provided in \autoref{fig:structure}.

\subsubsection{Kinesthetic Feedback for Compression Rate}
\par The system delivers phase-synchronized active haptic pacing based on real-time kinematic data. The control backend continuously estimates the user's compression rate, $f_{\mathrm{user}}$, by computing the moving average of the time intervals over the most recent $N$ compressions. Through pilot testing, we set the moving window size to $N = 5$:
\begin{equation}
f_{\mathrm{user}} = \frac{60\cdot N}{\sum_{i=1}^{N} \Delta T_i}
\label{eq:freq_calc}
\end{equation}
Here, $\Delta T_i$ denotes the time interval between two consecutive points of maximum compression depth.

\par When $f_{\mathrm{user}}$ deviates from the target range of 100-120 CPM, the system activates an active guidance mode. At each
control update, the activation state is defined as:

\begin{equation}
I_{\mathrm{rate}} =
\begin{cases}
1, & f_{\mathrm{user}} < 100~\mathrm{CPM}
     \ \text{or}\ 
     f_{\mathrm{user}} > 120~\mathrm{CPM}, \\
0, & 100~\mathrm{CPM}
     \leq f_{\mathrm{user}}
     \leq 120~\mathrm{CPM}.
\end{cases}
\label{eq:rate_activation}
\end{equation}

When $I_{\mathrm{rate}}$ switches from 0 to 1, the intervention is precisely timed to synchronize with the user's biomechanical phase~\cite{hove2014time}. Using the
laser displacement signal $z(t)$, the system detects velocity
zero-crossings and triggers a fixed active haptic pacing sequence at
110~CPM. Each pulse is initiated at the turning point corresponding
to maximum compression depth. At this moment, the proportional valve
outputs an instantaneous pressure peak, generating an upward physical
impulse that facilitates chest recoil. This periodic pacing force,
$F_{\mathrm{actuation}}(t)$, is modeled as a high-bandwidth square-wave
signal superimposed on the baseline resistance:

\begin{equation}
F_{\mathrm{actuation}}(t) =
F_{\mathrm{baseline}}
+ I_{\mathrm{rate}} \cdot \Delta F_{\mathrm{pulse}}
\sum_{k=0}^{\infty}
\Pi\left(
\frac{t - t_0 - k \cdot T_{\mathrm{target}}}
{\tau_{\mathrm{recoil}}}
\right),
\label{eq:force_pulse}
\end{equation}

where $T_{\mathrm{target}} = 0.545\mathrm{s}$ corresponds to a target
compression rate of 110~CPM, $\tau_{\mathrm{recoil}} = 100~\mathrm{ms}$
denotes the pulse duration used for rapid inflation, and
$\Delta F_{\mathrm{pulse}} = 200~\mathrm{N}$, determined through pilot
testing, is the additional instantaneous actuation force.
$\Pi(\cdot)$ denotes the rectangular pulse function. When
$I_{\mathrm{rate}}=0$, the pulse term is disabled and the system
maintains only the baseline resistance.

\subsubsection{Kinesthetic Guidance for Compression Depth}
\par For compression depth, the system provides active kinesthetic correction through a closed-loop variable-resistance mechanism, transforming the target depth range of 5-6~cm into a physically perceivable resistance boundary. The system continuously computes the moving average depth across the five most recent compressions, $\bar{d}_5(t)$:

\begin{equation}
\bar{d}_5(t)=\frac{1}{5}\sum_{i=0}^{4} d_{t-i},
\label{eq:depth_avg}
\end{equation}

where $d_{t-i}$ denotes the peak compression depth of the $(t-i)^{th}$ compression. Once $\bar{d}_5(t)$ falls outside the target interval $[d_{\min}, d_{\max}]$, with $d_{\min}=5$~cm and $d_{\max}=6$~cm, active depth guidance is triggered.

\par The controller regulates the pressure of the pneumatic chamber to adjust the effective chest resistance in real time. The baseline chest resistance is initialized as $F_{0}=100$~N, which serves as the default mechanical state during unguided practice. If the learner's recent compressions are consistently too shallow, the controller proportionally decreases the lower-chamber pressure, thereby reducing the resistance required to reach the target displacement. Conversely, if the average depth exceeds the target range, the controller increases the chamber pressure to create a physical barrier that guides the user back toward the desired range. The target actuation force $F_{\mathrm{target}}(t)$ is updated according to a proportional control law based on the depth error $e_d(t)$:

\begin{equation}
e_d(t)=\bar{d}_5(t)-d_{\mathrm{ref}},
\qquad
d_{\mathrm{ref}}=\frac{d_{\min}+d_{\max}}{2}=5.5~\mathrm{cm}
\label{eq:depth_error}
\end{equation}

\begin{equation}
F_{\mathrm{target}}(t)=
\mathrm{clip}\left(
F_{0}+K_d \cdot e_d(t),\,
F_{\min},\,
F_{\max}
\right)
\label{eq:depth_force}
\end{equation}

Here, $K_d$ is the depth control gain, and $F_{\min}$ and $F_{\max}$ constrain the resistance output within a safe range for smooth operation.

\par When the learner's average compression depth returns to the target interval, the assistive force is not removed abruptly. Instead, it is smoothly restored to the 100~N baseline through exponential decay:

\begin{equation}
F(t+1)=F(t)+\alpha \left(F_{0}-F(t)\right),
\label{eq:force_return}
\end{equation}

where $\alpha=0.1$ controls the rate at which the force returns to the baseline value. This soft-release mechanism avoids sudden discontinuities in chest-wall mechanics and operationalizes the system's principle of intervention on demand, ensuring that external scaffolding becomes active only under off-target conditions.

\subsection{Stage-Adaptive Fading Feedback}
\par To address \textbf{DG3}, \textit{Kinesthetic-CPR} implements a predefined three-stage fading schedule. This mechanism treats multimodal guidance as an externally provided scaffold with dynamically adjustable intensity and is designed to mitigate skill dependency that may arise from persistent external feedback. Internally, the system maintains a three-state intervention model that progressively shifts the responsibility for action calibration from the hardware system to the learner's proprioceptive system, achieved by reducing the richness of visual information and attenuating the physical intensity of kinesthetic intervention.

\par \textbf{Phase 1. Full Assistance.} In this phase, the system provides continuous multimodal intervention to accelerate the initial formation of motor patterns. The \textit{in-situ} MR interface remains persistently visible and presents immediate kinematic feedback for each compression. At the physical level, the closed-loop pneumatic system remains highly active and intervenes immediately through pneumatic pacing pulses and chest resistance reshaping whenever rhythm or depth deviates beyond the threshold. This state constitutes a comprehensive cognitive and physical scaffold during early learning.

\par \textbf{Phase 2. Attenuated Feedback.} In this phase, continuous scaffolding is converted into an on-demand conditional mechanism. The \textit{in-situ} visual interface enters a hidden mode and reappears only when the system detects substantial degradation in average rate or depth; once performance stabilizes, it fades out automatically. Meanwhile, the peak output of kinesthetic intervention is proportionally reduced, providing weaker rhythmic prompting and resistance correction. This state breaks the system's continuous dominance over action execution and encourages learners to maintain dynamic stability through internal perception during intervals between interventions.

\par \textbf{Phase 3. Minimal Assistance.} In this phase, informational redundancy is further stripped away, and feedback is downgraded from action guidance to simple error notification. The MR interface removes dynamic waveform and color mapping, retaining only textual labels for rate and depth that flash when violations occur, as shown in \autoref{fig:MR_interface}(d). On the kinesthetic side, the continuous resistance modulation algorithm is disabled, and the system delivers only a single transient pneumatic pulse as a physical reminder when error accumulation is detected. At this point, the full cognitive burden of planning actions and correcting errors is returned to the learner.

\section{Evaluation}
\par To answer \textbf{RQ3}, we conducted a controlled user study comprising two interconnected sub-studies. Study 1 employed a $2 \times 2$ between-subjects design to evaluate the independent and interactive effects of spatially aligned visual feedback and active kinesthetic guidance. Building on this, Study 2 introduced the stage-adaptive fading mechanism to assess the role of dynamically withdrawing external scaffolds in promoting skill internalization.

\subsection{General Method and Apparatus}

\subsubsection{Participants and Study Design}
\par We recruited 60 adult participants (P1-P60), primarily medical students recruited from a local medical school, all with normal upper-limb mobility and no prior systematic CPR training (\autoref{tab:appendix_participant-info-evaluation}). BLS certification is a prerequisite for clinical internships at this institution. During recruitment, we outlined the 30-day follow-up procedure and restricted enrollment to volunteers who confirmed both their willingness and availability to return. The participants were randomly and equally divided into five experimental groups ($N=12$ per group). The first four groups (G1-G4) constituted the $2 \times 2$ between-subjects design of Study 1, with the independent variables being the spatial location of visual feedback (ex-situ vs. in-situ) and kinesthetic assistance (without vs. with kinesthetic assistance). The fifth group (G5) constituted Study 2, utilizing the optimal multimodal feedback combination identified in Study 1, with the addition of the stage-adaptive fading strategy. Baseline comparability across groups was confirmed through pre-tests on age, gender, and prior CPR knowledge, revealing no significant statistical differences. Additionally, we invited five BLS-certified instructors (N1-N5) to experience the system and provide subsequent qualitative evaluations. All participants provided written informed consent. We explicitly assured them that their involvement, including the delayed test, was voluntary and would not affect course grades, BLS certification, examinations, or internship eligibility. To ensure complete retention, we scheduled all sessions in advance; consequently, all 60 enrolled participants returned for the delayed test and provided complete datasets.

\subsubsection{Apparatus and Task}
\par The experiment was conducted in a controlled lab environment. To maximize the fidelity of a real rescue posture, the custom pneumatic training manikin was placed on the floor, requiring participants to perform standard kneeling operations beside it. The driving cylinder and main control hardware of the pneumatic feedback module were concealed and soundproofed to eliminate additional attentional distractions caused by mechanical noise and equipment visibility. The experimental control backend recorded kinematic compression data in real time and streamed visual feedback rendering to either an external monitor or a Microsoft HoloLens 2 head-mounted display. To establish the ex-situ baseline, we adopted the screen-based configuration of our collaborating BLS instructors, wherein real-time compression metrics were presented on a front-facing television throughout practice. The experimental task strictly followed clinical guidelines: one complete cycle consisted of 30 chest compressions alternating with 2 rescue breaths, and each testing or training round required the continuous completion of 5 cycles.

\begin{figure}[h]
    \centering
    \includegraphics[width=\linewidth]{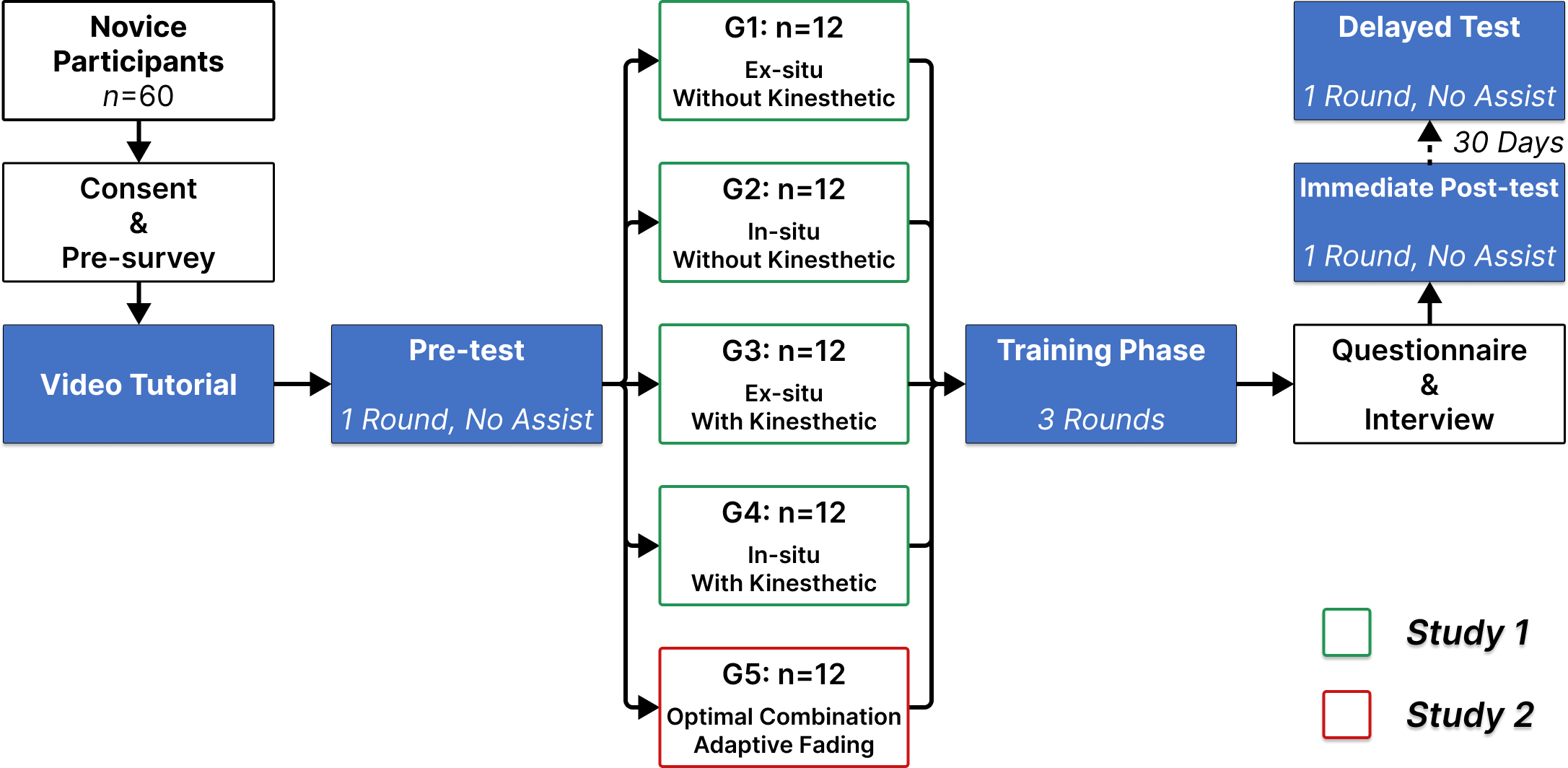}
    \caption{The workflow of our user study.}
    \Description{Flow diagram of the user study procedure for 60 novice participants. All participants first completed consent and a pre-survey, watched a video tutorial, and performed a one-round pre-test without assistance. They were then assigned to one of five groups with 12 participants each: G1, ex-situ without kinesthetic guidance; G2, in-situ without kinesthetic guidance; G3, ex-situ with kinesthetic guidance; G4, in-situ with kinesthetic guidance; and G5, the optimal combination with adaptive fading. After a three-round training phase, participants completed a questionnaire and interview, followed by an immediate post-test consisting of one round without assistance. A delayed test, also one round without assistance, was conducted 30 days later. The diagram distinguishes Study 1, which includes G1 to G4, from Study 2, which includes G5.}
    \label{fig:UserStudy}
\end{figure}

\subsubsection{Procedure}
\par The overall experimental procedure was standardized into four progressive phases: \textit{pre-test}, \textit{training phase}, \textit{immediate post-test}, and \textit{delayed test} (\autoref{fig:UserStudy}).
\begin{itemize}
    \item \textbf{Pre-test:} Participants first watched a standardized instructional video to align baseline procedural knowledge. Subsequently, they performed one independent round of CPR without any system assistance to record their initial performance.
    
    \item \textbf{Training Phase:} Participants underwent three rounds of training with specific interventions based on their assigned group. The four groups in Study 1 maintained a constant level of assistance intensity across the three rounds. Conversely, the stage-adaptive fading group in Study 2 followed a step-down strategy, gradually transitioning from full assistance to minimal assistance. To control for fatigue effects, a strict 1-minute rest period was enforced between consecutive training rounds.
    
    \item \textbf{Immediate Post-test:} After completing the training and filling out subjective questionnaires, participants returned to the unassisted state to complete another testing round, evaluating the immediate improvement in performance yielded by the system interventions.
    
    \item \textbf{Delayed Test:} Participants returned to the laboratory after 30 days to perform a final test under completely unassisted conditions, quantitatively assessing the long-term internalization and skill retention of the target motor schema.
\end{itemize}

\subsubsection{Measures and Data Analysis}
\par This study employed a mixed-methods approach to collect quantitative performance and qualitative perception data. Objective behavioral metrics included the average compression depth, average compression rate, and the corresponding success rates for these dimensions across all phases. Subjective evaluation metrics utilized the NASA-TLX~\cite{hart1988development} and System Usability Scale (SUS)~\cite{brooke1996sus} to quantify participants' cognitive load and system interaction experience. To eliminate statistical noise introduced by individual differences in motor aptitude, objective performance data were analyzed using Analysis of Covariance (ANCOVA)~\cite{rutherford2011anova}, with participants' pre-test scores incorporated as covariates. Subjective questionnaire data collected post-intervention were analyzed using Analysis of Variance (ANOVA)~\cite{st1989analysis}. Specifically, a two-way ANOVA was employed to assess the main and interaction effects of visual and kinesthetic feedback modalities on cognitive load and usability. Qualitative data were derived from the semi-structured interviews, employing thematic analysis to extract novices' perceptual characteristics regarding the feedback mechanisms and the professional instructors' evaluations of the system's pedagogical potential.

\subsection{Study 1: Effects of In-Situ Visualization and Kinesthetic Guidance}

\subsubsection{Immediate Post-Test Performance}
Study 1 examined the effects of visual feedback location (ex-situ vs.\ in-situ) and kinesthetic guidance (without vs.\ with guidance) on CPR performance immediately after training. Across the four fixed-support groups, the combination of in-situ visualization and kinesthetic guidance (G4) yielded the best immediate post-test performance. G4 achieved the highest depth accuracy (DA = 0.818) and rate accuracy (RA = 0.851), compared with G1 (DA = 0.595, RA = 0.668), G2 (DA = 0.617, RA = 0.678), and G3 (DA = 0.713, RA = 0.773). Its mean depth and mean rate (MD = 5.122 cm, MR = 103.252 CPM) were also closest to the midpoint of the target range (see Appendix \autoref{tab:training_results}).

ANCOVA showed significant main effects of both visual presentation and kinesthetic guidance on immediate post-test performance (see Appendix \autoref{tab:study1_ancova_rate}). For compression depth, there were significant effects of visual condition, $F(1,43)=32.75, p<.001, \eta_p^2=.432$, kinesthetic guidance, $F(1,43)=511.68, p<.001, \eta_p^2=.922$, and their interaction, $F(1,43)=64.25, p<.001, \eta_p^2=.599$. For compression rate, the main effects of visual condition, $F(1,43)=10.69, p=.002, \eta_p^2=.199$, kinesthetic guidance, $F(1,43)=233.98, p<.001, \eta_p^2=.845$, and their interaction, $F(1,43)=19.16, p<.001, \eta_p^2=.308$, were also significant. These results indicate that both in-situ visualization and kinesthetic guidance improved immediate post-test performance, with the combined condition producing the strongest short-term benefit.

\subsubsection{Delayed Retention Performance}
At the 30-day delayed post-test, G4 remained the best-performing group among the four fixed-support conditions. It again showed the highest depth accuracy (DA = 0.730) and rate accuracy (RA = 0.779), compared with G1 (DA = 0.497, RA = 0.573), G2 (DA = 0.542, RA = 0.603), and G3 (DA = 0.631, RA = 0.702). G4 also remained closest to the midpoint of the target range in terms of mean depth and mean rate (MD = 5.036 cm, MR = 101.135 CPM).

The ANCOVA results for delayed performance showed a weaker overall pattern than the immediate post-test. For compression depth, the main effects of visual condition, $F(1,43)=3.39, p=.073$, and kinesthetic guidance, $F(1,43)=0.22, p=.639$, were not significant, whereas their interaction remained significant, $F(1,43)=4.59, p=.038, \eta_p^2=.096$. For compression rate, neither the main effect of visual condition, $F(1,43)=3.00, p=.091$, nor kinesthetic guidance, $F(1,43)=2.65, p=.111$, reached significance, while the interaction effect remained significant, $F(1,43)=5.20, p=.028, \eta_p^2=.108$.

\subsubsection{Subjective Cognitive Load and System Usability}
Questionnaire results further supported the behavioral findings (see Appendix \autoref{fig:nasatlx}, \autoref{fig:sus}). For NASA-TLX, the four fixed-support groups differed most clearly in perceived performance and frustration. Among these conditions, G4 showed the highest self-rated performance and one of the lowest frustration levels, whereas G1 showed the lowest perceived performance and the highest frustration. By contrast, the remaining workload dimensions showed comparatively smaller between-group differences.

\subsection{Study 2: Effects of Stage-Adaptive Fading}

\subsubsection{Immediate Post-Test Performance}
Study 2 compared the best fixed-support condition identified in Study 1 (G4: in-situ visualization + kinesthetic guidance) with the stage-adaptive fading condition (G5: in-situ visualization + kinesthetic guidance + adaptive fading). At the immediate post-test, G4 showed better performance than G5 on both compression depth and compression rate. Specifically, G4 achieved higher depth accuracy (DA = 0.818) and rate accuracy (RA = 0.851) than G5 (DA = 0.785, RA = 0.826). G4 also showed slightly better mean performance, with mean depth and mean rate closer to the target range (G4: MD = 5.122 cm, MR = 103.252 CPM; G5: MD = 5.152 cm, MR = 102.356 CPM).

ANCOVA (see Appendix \autoref{tab:study2_ancova_rate}) confirmed this immediate advantage of fixed support. The group effect was significant for both depth, $F(1,21)=46.60, p<.001, \eta_p^2=.689$, and rate, $F(1,21)=10.13, p=.004, \eta_p^2=.325$. These results indicate that maintaining full multimodal support throughout training produced better short-term post-test performance than progressively fading assistance.

\subsubsection{Delayed Retention Performance}
At the 30-day delayed post-test, the pattern reversed. G5 outperformed G4 on both retained compression depth and retained compression rate. The fading group achieved higher depth accuracy (DA = 0.748 vs.\ 0.730) and rate accuracy (RA = 0.797 vs.\ 0.779). It also remained closer to the target range in terms of mean depth and mean rate (G5: MD = 5.125 cm, MR = 100.829 CPM; G4: MD = 5.036 cm, MR = 101.135 CPM).

This retention advantage was supported by ANCOVA. The fading condition significantly outperformed the fixed-support condition on delayed depth, $F(1,21)=21.39, p<.001, \eta_p^2=.505$, and delayed rate, $F(1,21)=40.08, p<.001, \eta_p^2=.656$. Together, these results show that although fixed multimodal support produced stronger immediate gains, stage-adaptive fading yielded better retained performance after external support was removed.

\section{Discussion}
Taken together, while in-situ visualization and kinesthetic guidance improved immediate performance, the adaptive fading results suggest that strong support alone is insufficient for long-term retention. These findings indicate that CPR training systems should be designed not only to optimize assisted performance but also to create conditions under which more stable bodily control can later be sustained without continuous support. We elaborate on this point through three key insights below.

\subsection{Key Insights}

\subsubsection{In-situ visualization preserves task-oriented coupling and reduces over-reliance on visual feedback}

\par Questionnaire results do not support a straightforward account in which the external-screen condition uniformly elevated perceived
workload. Instead, interview responses point to a potential difference in how participants related visual feedback to their bodily actions. For example, P5 described the external-screen experience as \textit{``more like playing a game of keeping the dashboard within the correct range than actually training.''} This characterization suggests that some learners treated the displayed indicators as a distinct performance target to be monitored, rather than as an informative resource integrated into their ongoing physical execution.

\par Complementing the behavioral advantage of the in-situ condition, this observation yields a design-level implication: feedback placement affects not only the cost of information access, but
also the functional role that information plays during practice. An external display may be construed as a detached performance target, whereas spatially aligned feedback can serve as an embedded corrective resource. By situating feedback precisely at the action site, in-situ visualization may help learners map visible errors onto bodily adjustments and maintain practice
anchored in physical execution. 

\par Given the absence of direct measures of gaze, attention allocation, or perceptual strategy, we frame this account as an interview-supported interpretation rather than a demonstrated mechanism. Future work incorporating eye tracking or systematic observation could more directly examine whether feedback location modulates attentional allocation during practice.

\subsubsection{Kinesthetic guidance shortens the motor correction loop}

Our results indicate that kinesthetic guidance did not simply add another feedback channel but provided qualitatively different support that improved immediate performance. In Study 1, its retention benefit appeared primarily in interaction with visual condition rather than as a standalone main effect, and questionnaire responses suggested lower perceived workload in the combined condition. While symbolic feedback informs learners of performance errors, kinesthetic guidance directly acts upon the ongoing movement, reducing the cognitive delay of decoding visual data into physical adjustments.

In early training phases, novices often lack an internal sensorimotor reference for correct compressions. Kinesthetic guidance shortens this correction loop by letting learners physically experience the target action pattern. This experiential calibration was reflected in one participant’s account: \textit{``During the first round of training, I found the manikin’s resistance very high... Just as I did not know how to adjust, I felt the resistance under my hands decrease slightly. Then I was able to control my compression depth within the correct range more easily, and I remembered that feeling''} (P28).

Moreover, our results showed that the advantage of kinesthetic guidance was more pronounced for compression depth than for rate. This pattern aligns with prior CPR research indicating that depth is harder to acquire than rate through indirect or technology-mediated training~\cite{ali2021cardiopulmonary}. Unlike rate, which can be easily paced by external audio-visual cues, depth requires learners to continuously regulate force against chest resistance and build a stable proprioceptive reference~\cite{johnson2016impact}. By transforming numerical depth targets into tangible mechanical boundaries, kinesthetic guidance grounds this bodily calibration in ways that symbolic feedback alone cannot.

\begin{figure}
    \centering
    \includegraphics[width=\linewidth]{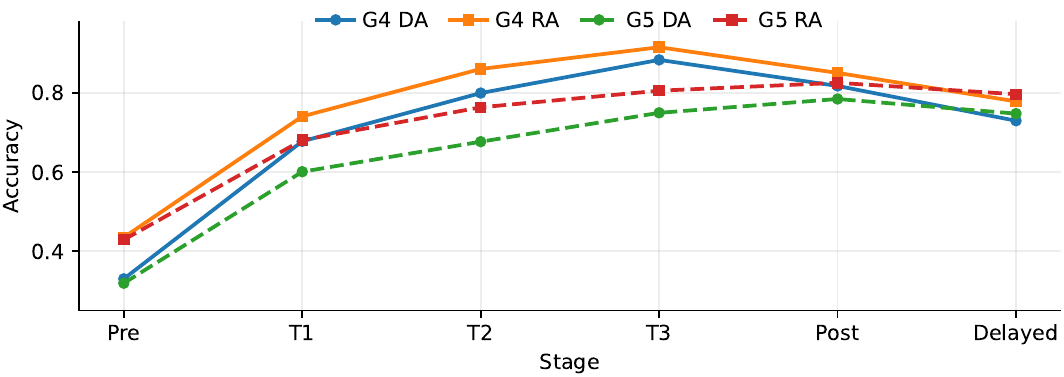}
    \caption{Learning curves of G4 and G5 across the study. G4 showed faster gains during training and better immediate post-test performance, whereas G5 exhibited better delayed retention in both depth accuracy and rate accuracy.}
    \Description{Line chart showing learning trajectories for two groups, G4 and G5, across six stages: Pre, T1, T2, T3, Post, and Delayed. The y-axis indicates accuracy. Four curves are plotted: G4 depth accuracy and rate accuracy, and G5 depth accuracy and rate accuracy. All measures increase substantially from Pre through the three training rounds. G4 reaches higher values than G5 during training and at the immediate post-test, with its highest performance around T3. At the delayed test, both G5 curves remain slightly higher than the corresponding G4 curves, indicating better retention for G5 in both depth and rate accuracy.}
    \label{fig:Curve}
\end{figure}

\subsubsection{Stage-adaptive fading may shift learners from system reliance to internal bodily reference}

\par Comparing the fixed-support condition (G4) with the stage-adaptive fading (G5) revealed divergent learning trajectories. G4 exhibited faster performance gains and superior immediate post-test outcomes, whereas G5 improved more gradually yet showed less decay after training, ultimately yielding better delayed post-test performance (\autoref{fig:Curve}).

\par By progressively reducing visual and kinesthetic scaffolding across the three training rounds, the stage-adaptive schedule required G5 participants to sustain or re-stabilize their performance with fewer external cues. Interview responses suggest that this process coincides with a shift in the cues participants relied upon. While those in the fixed-support groups (G1--G4) predominantly recalled the system's external cues, 75\% of G5 participants reported greater reliance on bodily sensations. Participants P53 and P59 further described a transition from system-dependent guidance to their own kinesthetic awareness. P53 characterized the later feedback as \textit{``a form of confirmation''} of their movements, while P59 described it as \textit{``just a subtle nudge to tell me I was not on the wrong track''}.

\par Taken together with the delayed post-test results, these accounts suggest that staged withdrawal may encourage learners to increasingly leverage bodily cues as external support diminishes, shifting the system's role from continuous action guidance to occasional performance verification. Given that we did not directly measure internalization or retention
strategies, we treat this explanation as an interview-informed interpretation rather than a definitively demonstrated mechanism.

\par This trade-off between immediate performance and long-term retention carries different
implications for CPR training versus assistance systems. Training systems aim to sustain performance after feedback is removed, making stage-adaptive fading particularly relevant for supporting durable skill retention. Real-time CPR assistance systems, in contrast, prioritize compression quality while feedback remains accessible, for which continuous support is likely more appropriate. Thus, the choice of feedback schedule should be guided by whether the system is intended to support skill retention or real-time execution.

\subsection{Design Implications for Psychomotor Skill Training}
Beyond the specific context of CPR, our findings suggest three broader implications for the design of psychomotor training systems that aim to support skill internalization. First, feedback should remain coupled with the site of action, rather than being presented as a detached interface to be monitored in parallel. For motor tasks that require sustained bodily attention, the goal of XR should not simply be to make information visible, but to preserve task-oriented perception and action. Second, symbolic feedback alone may be insufficient for skill dimensions that depend on force regulation, rhythm, or bodily coordination. In such cases, designers may need to complement visual guidance with bodily grounded forms of assistance that help learners directly experience the target action pattern. Third, training systems should be designed not only around how to provide assistance, but also around how to withdraw it. Our findings suggest that strong support is useful for early stabilization, whereas durable retention may depend on learners being gradually required to maintain performance under reduced guidance. Taken together, these implications highlight that psychomotor training should be understood not simply as an interface problem, but as the design of a staged transition from external scaffolding to internalized motor control.

\subsection{Limitations and Future Work}
While our findings highlight the potential of in-situ visualization, kinesthetic guidance, and adaptive fading for CPR training, several limitations remain.

\par \textit{Limitations in Assessing Skill Internalization}.  Motor internalization was inferred from immediate and 30-day delayed unassisted performance; these measures serve only as behavioral proxies and do not directly capture internalization. Future work could examine longer-term retention, transfer, and movement patterns.

\par \textit{Hardware Constraints of In-Situ Feedback}. The effectiveness of the MR feedback was constrained by the hardware. Some participants reported mild discomfort when using the HoloLens 2, and limited display brightness reduced the clarity of color-based cues. Therefore, the observed in-situ visualizations not only reflect the feedback design but also the capabilities of the current MR hardware.

\par \textit{Physical Demands of the Manikin}. Although the manikin was designed to match the physical response characteristics reported in prior literature, several experts noted that its chest felt harder than real chests. This may have imposed greater physical demands on participants, potentially accelerating fatigue and thereby affecting performance during the study.

\par \textit{Scope of CPR Metrics Evaluated}. Our evaluation focused only on compression depth and rate, while high-quality CPR also depends on full recoil, hand position, and minimizing interruptions. Whether the same design principles generalize to these additional metrics remains unclear.

\par \textit{Participant Population and Generalizability}. Our participants were primarily medical students without prior systematic CPR training. While they were novices to CPR, their medical education and certification-related motivation may distinguish them from community-based trainees, including older adults and individuals without a healthcare background. Thus, the present findings may not directly generalize to the broader community-training population. Future work should recruit more diverse community samples spanning a broad range of ages, educational backgrounds, physical abilities, and prior CPR experience to assess whether the observed effects of in-situ visualization, kinesthetic guidance, and stage-adaptive fading persist across different learner populations.

\section{Conclusion}
\par This paper presented \textit{Kinesthetic-CPR}, a multimodal CPR training system that combines in-situ MR visualization, active kinesthetic guidance, and adaptive feedback fading to support the internalization of psychomotor skill. Through a formative study, we identified three key design goals for CPR training: grounding feedback in the action space, coupling guidance with proprioceptive experience, and gradually reducing external scaffolding as learners improve. Our evaluation showed that spatially aligned visual feedback and kinesthetic guidance improved immediate performance, while adaptive fading better supported delayed retention after assistance was removed; questionnaire responses further indicated higher perceived performance and lower frustration under the combined multimodal condition. Together, these findings suggest that effective CPR training should be designed not only to improve performance during practice but also to help learners transition from external scaffolding to autonomous, embodied motor control.

\begin{acks}
We gratefully acknowledge the anonymous reviewers for their insightful feedback. This research was supported by the National Natural Science Foundation of China (No. 62372298), the Shanghai Engineering Research Center of Intelligent Vision and Imaging, the Shanghai Frontiers Science Center of Human-centered Artificial Intelligence (ShangHAI), and the Key Laboratory of Intelligent Perception and Human-Machine Collaboration (ShanghaiTech University), Ministry of Education.
\end{acks}

\bibliographystyle{ACM-Reference-Format}
\bibliography{AAReference}

@article{yang2012systematic,
  title={A systematic review of retention of adult advanced life support knowledge and skills in healthcare providers},
  author={Yang, Chih-Wei and Yen, Zui-Shen and McGowan, Jane E and Chen, Huiju Carrie and Chiang, Wen-Chu and Mancini, Mary E and Soar, Jasmeet and Lai, Mei-Shu and Ma, Matthew Huei-Ming},
  journal={Resuscitation},
  volume={83},
  number={9},
  pages={1055--1060},
  year={2012},
  publisher={Elsevier}
}

@article{michelland2023association,
  title={Association between basic life support and survival in sports-related sudden cardiac arrest: a meta-analysis},
  author={Michelland, Laurianne and Murad, Mohammad H and Bougouin, Wulfran and Van der Broek, Mark and Prokop, Larry J and Anys, Soraya and Perier, Marie-Cecile and Cariou, Alain and Empana, Jean Philippe and Marijon, Eloi and others},
  journal={European heart journal},
  volume={44},
  number={3},
  pages={180--192},
  year={2023},
  publisher={Oxford University Press US}
}

@article{perkins2015european,
  title={European Resuscitation Council Guidelines for Resuscitation 2015: Section 2. Adult basic life support and automated external defibrillation},
  author={Perkins, Gavin D and Handley, Anthony J and Koster, Rudolph W and Castr{\'e}n, Maaret and Smyth, Michael A and Olasveengen, Theresa and Monsieurs, Koenraad G and Raffay, Violetta and Gr{\"a}sner, Jan-Thorsten and Wenzel, Volker and others},
  journal={Resuscitation},
  volume={95},
  pages={81--99},
  year={2015},
  publisher={Elsevier}
}

@article{marijon2023lancet,
  title={The Lancet Commission to reduce the global burden of sudden cardiac death: a call for multidisciplinary action},
  author={Marijon, Eloi and Narayanan, Kumar and Smith, Karen and Barra, Sergio and Basso, Cristina and Blom, Marieke T and Crotti, Lia and d'Avila, Andre and Deo, Rajat and Dumas, Florence and others},
  journal={The Lancet},
  volume={402},
  number={10405},
  pages={883--936},
  year={2023},
  publisher={Elsevier}
}

@article{baldi2017real,
  title={Real-time visual feedback during training improves laypersons’ CPR quality: a randomized controlled manikin study},
  author={Baldi, Enrico and Cornara, Stefano and Contri, Enrico and Epis, Francesco and Fina, Dario and Zelaschi, Beatrice and Dossena, Cinzia and Fichtner, Ferdinando and Tonani, Michela and Di Maggio, Marzia and others},
  journal={Canadian Journal of Emergency Medicine},
  volume={19},
  number={6},
  pages={480--487},
  year={2017},
  publisher={Cambridge University Press}
}

@article{wagner2022visual,
  title={Visual attention during pediatric resuscitation with feedback devices: a randomized simulation study},
  author={Wagner, Michael and Gr{\"o}pel, Peter and Eibensteiner, Felix and Kessler, Lisa and Bibl, Katharina and Gross, Isabel T and Berger, Angelika and Cardona, Francesco S},
  journal={Pediatric Research},
  volume={91},
  number={7},
  pages={1762--1768},
  year={2022},
  publisher={Nature Publishing Group US New York}
}

@article{liu2025electrotactile,
author = {liu, zhiming and Meng, Yibo and xiao, yiqi and tian, sunan and cai, xudong and zhang, Guoli and su, hankun and lei, zhihao and hu, zhefang},
year = {2025},
month = {07},
pages = {},
title = {Electrotactile, Vibrotactile, and Thermal Feedback in a Wearable CPR Training Game: Comparative Effectiveness Study (Preprint)},
doi = {10.2196/preprints.80768}
}

@article{schmidt1989summary,
  title={Summary knowledge of results for skill acquisition: support for the guidance hypothesis.},
  author={Schmidt, Richard A and Young, Douglas E and Swinnen, Stephan and Shapiro, Diane C},
  journal={Journal of Experimental Psychology: Learning, Memory, and Cognition},
  volume={15},
  number={2},
  pages={352},
  year={1989},
  publisher={American Psychological Association}
}

@book{fitts1967human,
  title={Human performance.},
  author={Fitts, Paul M and Posner, Michael I},
  year={1967},
  publisher={Brooks/Cole}
}

@article{yeung2009use,
  title={The use of CPR feedback/prompt devices during training and CPR performance: a systematic review},
  author={Yeung, Joyce and Meeks, Reylon and Edelson, Dana and Gao, Fang and Soar, Jasmeet and Perkins, Gavin D},
  journal={Resuscitation},
  volume={80},
  number={7},
  pages={743--751},
  year={2009},
  publisher={Elsevier}
}

@inproceedings{sense2016interactions,
  title={Interactions of declarative and procedural memory in real-life tasks: validating CPR as a new paradigm},
  author={Sense, Florian and Maass, Sarah and van Rijn, Hedderik},
  booktitle={Proceedings of the 14th International Conference on Cognitive Modeling},
  year={2016}
}

@article{burkhardt2014effect,
  title={Effect of prior cardiopulmonary resuscitation knowledge on compression performance by hospital providers},
  author={Burkhardt, Joshua N and Glick, Joshua E and Terndrup, Thomas E},
  journal={Western Journal of Emergency Medicine},
  volume={15},
  number={4},
  pages={404},
  year={2014}
}

@article{schmidt1975schema,
  title={A schema theory of discrete motor skill learning.},
  author={Schmidt, Richard A},
  journal={Psychological review},
  volume={82},
  number={4},
  pages={225},
  year={1975},
  publisher={American Psychological Association}
}

@article{white2017measuring,
  title={Measuring the effectiveness of a novel CPRcard™ feedback device during simulated chest compressions by non-healthcare workers},
  author={White, Alexander E and Ng, Han Xian and Ng, Wai Yee and Ng, Eileen Kai Xin and Fook-Chong, Stephanie and Kua, Phek Hui Jade and Ong, Marcus Eng Hock},
  journal={Singapore Medical Journal},
  volume={58},
  number={7},
  pages={438},
  year={2017}
}

@inproceedings{gruenerbl2018training,
  title={Training CPR with a wearable real time feedback system},
  author={Gruenerbl, Agnes and Javaheri, Hamraz and Monger, Eloise and Gobbi, Mary and Lukowicz, Paul},
  booktitle={Proceedings of the 2018 ACM International Symposium on Wearable Computers},
  pages={44--47},
  year={2018}
}

@inproceedings{di2020real,
  title={Real-time multimodal feedback with the CPR tutor},
  author={Di Mitri, Daniele and Schneider, Jan and Trebing, Kevin and Sopka, Sasa and Specht, Marcus and Drachsler, Hendrik},
  booktitle={International conference on artificial intelligence in education},
  pages={141--152},
  year={2020},
  organization={Springer}
}

@mastersthesis{himane2025stayin,
  title={Stayin' Alive! - Implementing Haptic Feedback For CPR Training in Virtual Reality},
  author={Himane, Sridhar and Vella, Arik},
  year={2025},
  school={Ume{\aa} University},
  type={Independent thesis Advanced level (degree of Master (Two Years))}
}

@article{chandler1992split,
  title={The split-attention effect as a factor in the design of instruction},
  author={Chandler, Paul and Sweller, John},
  journal={British Journal of Educational Psychology},
  volume={62},
  number={2},
  pages={233--246},
  year={1992},
  publisher={Wiley Online Library}
}

@inproceedings{park2013projected,
  title={Projected AR-based interactive CPR simulator},
  author={Park, Nohyoung and Kwon, Yeram and Lee, Sungwon and Woo, Woontack and Jeong, Jihoon},
  booktitle={International Conference on Virtual, Augmented and Mixed Reality},
  pages={83--89},
  year={2013},
  organization={Springer}
}

@article{lim2021rapid,
  title={Rapid degradation of psychomotor memory causes poor quality chest compressions in frequent cardiopulmonary resuscitation providers and feedback devices can only help to a limited degree: A crossover simulation study},
  author={Lim, Wan Yen and Ong, John and Ong, Sharon and Teo, LM and Fook-Chong, S and Ho, VK},
  journal={Medicine},
  volume={100},
  number={8},
  pages={e23927},
  year={2021},
  publisher={LWW}
}

@book{schmidt2008motor,
  title={Motor learning and performance: A situation-based learning approach},
  author={Schmidt, Richard A and Wrisberg, Craig A},
  year={2008},
  publisher={Human kinetics}
}

@incollection{adams1976issues,
  title={Issues for a closed-loop theory of motor learning},
  author={Adams, Jack A},
  booktitle={Motor control},
  pages={87--107},
  year={1976},
  publisher={Elsevier}
}

@article{zhang2026seconds,
  title={When Seconds Count: Designing Real-Time VR Interventions for Stress Inoculation Training in Novice Physicians},
  author={Zhang, Shuhao and Dong, Jiahe and Wang, Haoran and Jiang, Chang and Li, Quan},
  journal={arXiv preprint arXiv:2601.17458},
  year={2026}
}

@incollection{ohshima2023mr,
  title={MR BLS Trainer: A physical mixed reality CPR+ AED rescue simulator},
  author={Ohshima, Toshikazu and Matsui, Saina and Yamane, Mizuki and Ling, Yali and Muroi, Katsuhito and Sakai, Chihiro},
  booktitle={SIGGRAPH Asia 2023 XR},
  pages={1--2},
  year={2023}
}

@inproceedings{durai2019effect,
  title={The effect of audio and visual modality based cpr skill training with haptics feedback in vr},
  author={Durai, Varun SI and Arjunan, Raj and Manivannan, M},
  booktitle={2019 IEEE Conference on Virtual Reality and 3D User Interfaces (VR)},
  pages={910--911},
  year={2019},
  organization={IEEE}
}

@article{sarac2026multimodal,
  title={Multimodal Feedback in Automated Training Systems for Cardiopulmonary Resuscitation: A Review},
  author={Sarac, Mine and Cetin, Sevval and Aslan, Baran and Tas, Ali and Stroppa, Fabio},
  journal={IEEE Transactions on Human-Machine Systems},
  year={2026},
  publisher={IEEE}
}

@article{nysaether2008manikins,
  title={Manikins with human-like chest properties—a new tool for chest compression research},
  author={Nysaether, Jon B and Dorph, Elizabeth and Rafoss, Ivan and Steen, Petter A},
  journal={IEEE Transactions on Biomedical Engineering},
  volume={55},
  number={11},
  pages={2643--2650},
  year={2008},
  publisher={IEEE}
}

@inproceedings{xie2009simulator,
  title={A simulator of human chest that simulated force-sternal displacement relationship during cardiopulmonary resuscitation},
  author={Xie, Xinwu and Tian, Feng and Sun, Qiuming and Wang, Zheng and Aijuan, Ni and Hu, Mingxi},
  booktitle={2009 3rd International Conference on Bioinformatics and Biomedical Engineering},
  pages={1--4},
  year={2009},
  organization={IEEE}
}

@inproceedings{stanley2012recreating,
  title={Recreating the feel of the human chest in a CPR manikin via programmable pneumatic damping},
  author={Stanley, Andrew A and Healey, Simon K and Maltese, Matthew R and Kuchenbecker, Katherine J},
  booktitle={2012 IEEE Haptics Symposium (HAPTICS)},
  pages={37--44},
  year={2012},
  organization={IEEE}
}

@article{thielen2017innovative,
  title={An innovative design for cardiopulmonary resuscitation manikins based on a human-like thorax and embedded flow sensors},
  author={Thielen, Mark and Joshi, Rohan and Delbressine, Frank and Bambang Oetomo, Sidarto and Feijs, Loe},
  journal={Proceedings of the Institution of Mechanical Engineers, Part H: Journal of Engineering in Medicine},
  volume={231},
  number={3},
  pages={243--249},
  year={2017},
  publisher={SAGE Publications Sage UK: London, England}
}

@article{semeraro2019back,
  title={Back to reality: a new blended pilot course of basic life support with virtual reality},
  author={Semeraro, Federico and Ristagno, Giuseppe and Giulini, Gabriele and Kayal, Jihan Samira and Cavallo, Piergiorgio and Farabegoli, Lucia and Tucci, Riccardo and Scelsi, Silvia and Grieco, Niccol{\`o} Brenno and Scapigliati, Andrea},
  journal={Resuscitation},
  volume={138},
  pages={18--19},
  year={2019},
  publisher={Elsevier}
}

@article{meaney2013cardiopulmonary,
  title={Cardiopulmonary resuscitation quality: improving cardiac resuscitation outcomes both inside and outside the hospital: a consensus statement from the American Heart Association},
  author={Meaney, Peter A and Bobrow, Bentley J and Mancini, Mary E and Christenson, Jim and de Caen, Allan R and Bhanji, Farhan and Abella, Benjamin S and Kleinman, Monica E and Edelson, Dana P and Berg, Robert A and others},
  journal={Circulation},
  volume={128},
  number={4},
  pages={417--435},
  year={2013},
  publisher={Lippincott Williams \& Wilkins Hagerstown, MD}
}

@article{gruben1993sternal,
  title={Sternal force-displacement relationship during cardiopulmonary resuscitation},
  author={Gruben, Kreg G and Guerci, AD and Halperin, HR and Popel, AS and Tsitlik, JE},
  year={1993}
}

@article{tanaka2019effect,
  title={Effect of real-time visual feedback device ‘Quality Cardiopulmonary Resuscitation (QCPR) Classroom’with a metronome sound on layperson CPR training in Japan: a cluster randomized control trial},
  author={Tanaka, Shota and Tsukigase, Kyoko and Hara, Takahiro and Sagisaka, Ryo and Myklebust, Helge and Birkenes, Tonje Soraas and Takahashi, Hiroyuki and Iwata, Ayana and Kidokoro, Yutaro and Yamada, Momoyo and others},
  journal={BMJ open},
  volume={9},
  number={6},
  pages={e026140},
  year={2019},
  publisher={British Medical Journal Publishing Group}
}

@article{buchner2022impact,
  title={The impact of augmented reality on cognitive load and performance: A systematic review},
  author={Buchner, Josef and Buntins, Katja and Kerres, Michael},
  journal={Journal of Computer Assisted Learning},
  volume={38},
  number={1},
  pages={285--303},
  year={2022},
  publisher={Wiley Online Library}
}

@article{johnson2012eye,
  title={An eye movement analysis of the spatial contiguity effect in multimedia learning.},
  author={Johnson, Cheryl I and Mayer, Richard E},
  journal={Journal of Experimental Psychology: Applied},
  volume={18},
  number={2},
  pages={178},
  year={2012},
  publisher={American Psychological Association}
}

@inproceedings{johnson2018holocpr,
  title={Holocpr: Designing and evaluating a mixed reality interface for time-critical emergencies},
  author={Johnson, Janet G and Rodrigues, Danilo Gasques and Gubbala, Madhuri and Weibel, Nadir},
  booktitle={Proceedings of the 12th EAI International Conference on Pervasive Computing Technologies for Healthcare},
  pages={67--76},
  year={2018}
}

@article{odegaard2007chest,
  title={Chest compressions by ambulance personnel on chests with variable stiffness: abilities and attitudes},
  author={{\O}degaard, Silje and Kramer-Johansen, Jo and Bromley, Allan and Myklebust, Helge and Nys{\ae}ther, Jon and Wik, Lars and Steen, Petter Andreas},
  journal={Resuscitation},
  volume={74},
  number={1},
  pages={127--134},
  year={2007},
  publisher={Elsevier}
}

@inproceedings{grindlay2008haptic,
  title={Haptic guidance benefits musical motor learning},
  author={Grindlay, Graham},
  booktitle={2008 Symposium on Haptic Interfaces for Virtual Environment and Teleoperator Systems},
  pages={397--404},
  year={2008},
  organization={IEEE}
}

@article{gutierrez2024immersive,
  title={Immersive haptic simulation for training nurses in emergency medical procedures: Evaluation of the needle decompression procedure: HMD and haptics versus mannequin},
  author={Guti{\'e}rrez-Fern{\'a}ndez, Alexis and Fern{\'a}ndez-Llamas, Camino and V{\'a}zquez-Casares, Ana M and Mauriz, Elba and Riego-del-Castillo, Virginia and John, Nigel W},
  journal={The Visual Computer},
  volume={40},
  number={11},
  pages={7527--7537},
  year={2024},
  publisher={Springer}
}

@article{gomez2025simulation,
  title={Simulation training with haptic feedback of instrument vibrations reduces resident workload during live robot-assisted sleeve gastrectomy},
  author={Gomez, Ernest D and Husin, Haliza Mat and Dumon, Kristoffel R and Williams, Noel N and Kuchenbecker, Katherine J},
  journal={Surgical Endoscopy},
  volume={39},
  number={3},
  pages={1523--1535},
  year={2025},
  publisher={Springer}
}

@article{aman2015effectiveness,
  title={The effectiveness of proprioceptive training for improving motor function: a systematic review},
  author={Aman, Joshua E and Elangovan, Naveen and Yeh, I-Ling and Konczak, J{\"u}rgen},
  journal={Frontiers in human neuroscience},
  volume={8},
  pages={1075},
  year={2015},
  publisher={Frontiers Media SA}
}

@article{sigrist2013augmented,
  title={Augmented visual, auditory, haptic, and multimodal feedback in motor learning: a review},
  author={Sigrist, Roland and Rauter, Georg and Riener, Robert and Wolf, Peter},
  journal={Psychonomic bulletin \& review},
  volume={20},
  number={1},
  pages={21--53},
  year={2013},
  publisher={Springer}
}

@inproceedings{huegel2010progressive,
  title={Progressive haptic and visual guidance for training in a virtual dynamic task},
  author={Huegel, Joel C and O'Malley, Marcia K},
  booktitle={2010 IEEE Haptics Symposium},
  pages={343--350},
  year={2010},
  organization={IEEE}
}

@article{braun2006using,
  title={Using thematic analysis in psychology},
  author={Braun, Virginia and Clarke, Victoria},
  journal={Qualitative research in psychology},
  volume={3},
  number={2},
  pages={77--101},
  year={2006},
  publisher={Taylor \& Francis}
}

@article{willett2016embedded,
  title={Embedded data representations},
  author={Willett, Wesley and Jansen, Yvonne and Dragicevic, Pierre},
  journal={IEEE transactions on visualization and computer graphics},
  volume={23},
  number={1},
  pages={461--470},
  year={2016},
  publisher={IEEE}
}

@incollection{marriott2018immersive,
  title={Immersive analytics: Time to reconsider the value of 3d for information visualisation},
  author={Marriott, Kim and Chen, Jian and Hlawatsch, Marcel and Itoh, Takayuki and Nacenta, Miguel A and Reina, Guido and Stuerzlinger, Wolfgang},
  booktitle={Immersive analytics},
  pages={25--55},
  year={2018},
  publisher={Springer}
}

@inproceedings{munzner2025visualization,
  title={Visualization analysis and design},
  author={Munzner, Tamara},
  booktitle={Proceedings of the Special Interest Group on Computer Graphics and Interactive Techniques Conference Courses},
  pages={1--2},
  year={2025}
}

@book{ware2019information,
  title={Information visualization: perception for design},
  author={Ware, Colin},
  year={2019},
  publisher={Morgan Kaufmann}
}

@inproceedings{lu2020glanceable,
  title={Glanceable ar: Evaluating information access methods for head-worn augmented reality},
  author={Lu, Feiyu and Davari, Shakiba and Lisle, Lee and Li, Yuan and Bowman, Doug A},
  booktitle={2020 IEEE conference on virtual reality and 3D user interfaces (VR)},
  pages={930--939},
  year={2020},
  organization={IEEE}
}

@incollection{hart1988development,
  title={Development of NASA-TLX (Task Load Index): Results of empirical and theoretical research},
  author={Hart, Sandra G and Staveland, Lowell E},
  booktitle={Advances in psychology},
  volume={52},
  pages={139--183},
  year={1988},
  publisher={Elsevier}
}

@article{brooke1996sus,
  title={SUS-A quick and dirty usability scale},
  author={Brooke, John and others},
  journal={Usability evaluation in industry},
  volume={189},
  number={194},
  pages={4--7},
  year={1996},
  publisher={London, England}
}

@article{st1989analysis,
  title={Analysis of variance (ANOVA)},
  author={St, Lars and Wold, Svante and others},
  journal={Chemometrics and intelligent laboratory systems},
  volume={6},
  number={4},
  pages={259--272},
  year={1989},
  publisher={Elsevier}
}

@book{rutherford2011anova,
  title={ANOVA and ANCOVA: a GLM approach},
  author={Rutherford, Andrew},
  volume={658},
  year={2011},
  publisher={Wiley Online Library}
}

@article{hamam2013effect,
  title={Effect of kinesthetic and tactile haptic feedback on the quality of experience of edutainment applications},
  author={Hamam, Abdelwahab and Eid, Mohamad and El Saddik, Abdulmotaleb},
  journal={Multimedia tools and applications},
  volume={67},
  number={2},
  pages={455--472},
  year={2013},
  publisher={Springer}
}

@article{hove2014time,
  title={The time course of phase correction: a kinematic investigation of motor adjustment to timing perturbations during sensorimotor synchronization.},
  author={Hove, Michael J and Balasubramaniam, Ramesh and Keller, Peter E},
  journal={Journal of Experimental Psychology: Human Perception and Performance},
  volume={40},
  number={6},
  pages={2243},
  year={2014},
  publisher={American Psychological Association}
}

@article{schipper2025technological,
  title={Technological innovations in layperson CPR education--a scoping review},
  author={Schipper, Abigail E and Sloane, Charles SM and Shimelis, Lydia B and Kim, Ryan T},
  journal={Resuscitation Plus},
  volume={23},
  pages={100924},
  year={2025},
  publisher={Elsevier}
}

@article{ali2021cardiopulmonary,
  title={Cardiopulmonary resuscitation (CPR) training strategies in the times of COVID-19: a systematic literature review comparing different training methodologies},
  author={Ali, Daniyal Mansoor and Hisam, Butool and Shaukat, Natasha and Baig, Noor and Ong, Marcus Eng Hock and Epstein, Jonathan L and Goralnick, Eric and Kivela, Paul D and McNally, Bryan and Razzak, Junaid},
  journal={Scandinavian journal of trauma, resuscitation and emergency medicine},
  volume={29},
  number={1},
  pages={53},
  year={2021},
  publisher={Springer}
}

@article{johnson2016impact,
  title={The impact of quantitative feedback on the performance of chest compression by basic life support trained clinical staff},
  author={Johnson, Matthew and Peat, Amanda and Boyd, Leanne and Warren, Tanya and Eastwood, Kathryn and Smith, Gavin},
  journal={Nurse Education Today},
  volume={45},
  pages={163--166},
  year={2016},
  publisher={Elsevier}
}

@misc{smc_itv3050,
  author       = {{SMC Corporation}},
  title        = {{Electro-Pneumatic Regulator, ITV3000 Series (Model ITV3050)}},
  year         = {n.d.},
  howpublished = {\url{https://www.smcworld.com/}},
  note         = {Accessed 2026-04-01}
}

@misc{microsoft_hololens2,
  author       = {{Microsoft}},
  title        = {{HoloLens 2 Hardware}},
  year         = {2025},
  howpublished = {\url{https://learn.microsoft.com/zh-tw/hololens/hololens2-hardware}},
  note         = {Accessed 2026-04-01}
}

@inproceedings{anderson2013youmove,
  author    = {Anderson, Fraser and Grossman, Tovi and
               Matejka, Justin and Fitzmaurice, George},
  title     = {{YouMove}: Enhancing Movement Training with an
               Augmented Reality Mirror},
  booktitle = {Proceedings of the 26th Annual ACM Symposium on
               User Interface Software and Technology},
  year      = {2013},
  pages     = {311--320},
  doi       = {10.1145/2501988.250204}
}

@article{leary2020pilot,
  author  = {Leary, Marion and McGovern, Shaun K. and Balian, Steve
             and Abella, Benjamin S. and Blewer, Audrey L.},
  title   = {A Pilot Study of CPR Quality Comparing an Augmented
             Reality Application vs. a Standard Audio-Visual
             Feedback Manikin},
  journal = {Frontiers in Digital Health},
  volume  = {2},
  year    = {2020},
  pages   = {1},
  doi     = {10.3389/fdgth.2020.00001}
}

@inproceedings{iannucci2023arrow,
  author    = {Iannucci, Elena and Chen, Zhutian and Armeni, Iro and
               Pollefeys, Marc and Pfister, Hanspeter and
               Beyer, Johanna},
  title     = {{ARrow}: A Real-Time AR Rowing Coach},
  booktitle = {EuroVis 2023 -- Short Papers},
  year      = {2023},
  publisher = {The Eurographics Association},
  doi       = {10.2312/evs.20231046}
}

@inproceedings{ihara2025video2mr,
  author    = {Ihara, Keiichi and Monteiro, Kyzyl and Faridan, Mehrad
               and Kazi, Rubaiat Habib and Suzuki, Ryo},
  title     = {{Video2MR}: Automatically Generating Mixed Reality
               3D Instructions by Augmenting Extracted Motion from
               2D Videos},
  booktitle = {Proceedings of the 30th International Conference on
               Intelligent User Interfaces},
  year      = {2025},
  doi       = {10.1145/3708359.3712159}
}

@inproceedings{lee2026vistar,
  author    = {Lee, Chunggi and Saiki, Hayato and Lin, Tica and
               Ikeda, Eiji and Suzuki, Kenji and Zhu-Tian, Chen and
               Pfister, Hanspeter},
  title     = {{ViSTAR}: Virtual Skill Training with Augmented
               Reality with 3D Avatars and LLM Coaching Agent},
  booktitle = {Proceedings of the 2026 CHI Conference on Human
               Factors in Computing Systems},
  year      = {2026},
  doi       = {10.1145/3772318.3790634}
}


\appendix

\section{Participant Demographics}

\begin{table}[H]
    \centering
    \caption{Participant Demographics in Formative Study.}
    \Description{Table listing formative study participant demographics for 14 participants, including ID, age, gender, and group. The sample comprises 8 novices labeled T1 to T8 and 6 instructors labeled I1 to I6. Novices are 22 to 24 years old and include 3 men and 5 women. Instructors are 33 to 41 years old and include 3 men and 3 women. The table shows that the formative study involved a younger novice group and an older instructor group with mixed gender representation in both groups.}
    \label{tab:appendix_participant-info-formative}
    \begin{tabular}{@{}llll@{}}
        \toprule
        ID & Age & Gender & Group \\ \midrule
        T1 & 23 & M & Novice \\
        T2 & 24 & F & Novice \\
        T3 & 23 & M & Novice \\
        T4 & 23 & F & Novice \\
        T5 & 24 & M & Novice \\
        T6 & 22 & F & Novice \\
        T7 & 23 & F & Novice \\
        T8 & 22 & F & Novice \\
        I1 & 33 & M & Instructor \\
        I2 & 38 & F & Instructor \\
        I3 & 41 & F & Instructor \\
        I4 & 39 & M & Instructor \\
        I5 & 37 & M & Instructor \\
        I6 & 38 & F & Instructor \\ \bottomrule
    \end{tabular}
\end{table}

\begin{table}[H]
    \centering
    \footnotesize
    \setlength{\tabcolsep}{3pt}
    \caption{Participant Demographics in Evaluation.}
    \Description{Table listing evaluation-study participant demographics for 60 participants, including ID, age, gender, and assigned group. Participants are labeled P1 to P60 and are evenly distributed across five groups, G1 to G5, with 12 participants per group. Ages range from 18 to 23 years. Across the full sample, there are 32 men and 28 women. The table primarily documents the individual participant assignments and shows that the five evaluation groups are similarly sized with comparable age and gender composition.}
    \label{tab:appendix_participant-info-evaluation}
    \begin{minipage}[t]{0.48\columnwidth}
        \centering
        \begin{tabular}{@{}llll@{}}
            \toprule
            ID & Age & Gender & Group \\ \midrule
            P1  & 21 & M & G1 \\
            P2  & 20 & F & G1 \\
            P3  & 23 & M & G1 \\
            P4  & 19 & F & G1 \\
            P5  & 19 & M & G1 \\
            P6  & 22 & F & G1 \\
            P7  & 19 & M & G1 \\
            P8  & 20 & M & G1 \\
            P9  & 18 & M & G1 \\
            P10 & 23 & M & G1 \\
            P11 & 21 & F & G1 \\
            P12 & 20 & F & G1 \\
            P13 & 20 & M & G2 \\
            P14 & 20 & M & G2 \\
            P15 & 22 & F & G2 \\
            P16 & 20 & M & G2 \\
            P17 & 20 & M & G2 \\
            P18 & 23 & F & G2 \\
            P19 & 19 & M & G2 \\
            P20 & 23 & F & G2 \\
            P21 & 23 & M & G2 \\
            P22 & 19 & F & G2 \\
            P23 & 19 & F & G2 \\
            P24 & 23 & M & G2 \\
            P25 & 19 & F & G3 \\
            P26 & 22 & F & G3 \\
            P27 & 19 & M & G3 \\
            P28 & 23 & F & G3 \\
            P29 & 20 & M & G3 \\
            P30 & 20 & M & G3 \\ \bottomrule
        \end{tabular}
    \end{minipage}\hfill
    \begin{minipage}[t]{0.48\columnwidth}
        \centering
        \begin{tabular}{@{}llll@{}}
            \toprule
            ID & Age & Gender & Group \\ \midrule
            P31 & 23 & F & G3 \\
            P32 & 23 & F & G3 \\
            P33 & 20 & M & G3 \\
            P34 & 19 & M & G3 \\
            P35 & 22 & M & G3 \\
            P36 & 20 & F & G3 \\
            P37 & 22 & M & G4 \\
            P38 & 23 & M & G4 \\
            P39 & 20 & F & G4 \\
            P40 & 23 & F & G4 \\
            P41 & 20 & M & G4 \\
            P42 & 21 & F & G4 \\
            P43 & 21 & M & G4 \\
            P44 & 21 & M & G4 \\
            P45 & 21 & F & G4 \\
            P46 & 23 & F & G4 \\
            P47 & 23 & M & G4 \\
            P48 & 22 & F & G4 \\
            P49 & 19 & M & G5 \\
            P50 & 22 & M & G5 \\
            P51 & 19 & F & G5 \\
            P52 & 21 & F & G5 \\
            P53 & 20 & F & G5 \\
            P54 & 20 & M & G5 \\
            P55 & 20 & M & G5 \\
            P56 & 20 & F & G5 \\
            P57 & 22 & M & G5 \\
            P58 & 22 & F & G5 \\
            P59 & 19 & F & G5 \\
            P60 & 22 & M & G5 \\ \bottomrule
        \end{tabular}
    \end{minipage}
\end{table}

\section{Numerical results of the user study}

\begin{table*}[htbp]
\centering
\caption{Performance of Each Group Across the Pretest and the First Two Training Rounds}
\Description{Two-part table reporting group performance across six phases: Pretest, Train 1, Train 2, Train 3, Posttest, and Delayed Posttest. For each of five groups, G1 to G5, the table lists mean depth, mean rate, depth accuracy, and rate accuracy. Pretest values are similar across groups, with mean depth around 4.4 to 4.6, mean rate around 96.7 to 97.3, depth accuracy around 0.30 to 0.33, and rate accuracy around 0.41 to 0.44. Across the three training rounds, all groups improve in both depth accuracy and rate accuracy. G4 achieves the highest accuracy during Train 1, Train 2, Train 3, and the immediate Posttest, reaching 0.884 depth accuracy and 0.916 rate accuracy at Train 3, and 0.818 depth accuracy and 0.851 rate accuracy at Posttest. At the Delayed Posttest, G5 has the highest retained accuracy, with 0.748 depth accuracy and 0.797 rate accuracy, slightly above G4. Mean depth generally increases toward the target range during training and remains near 5.0 in later phases, while mean rate rises during training and decreases somewhat at posttest and delayed posttest. Overall, the table shows broad learning gains for all groups, strongest immediate performance for G4, and best delayed retention for G5.}
\label{tab:training_results}

\resizebox{\textwidth}{!}{%
\begin{tabular}{lcccccccccccc}
\toprule
\multirow{2}{*}{Group}
& \multicolumn{4}{c}{Pretest}
& \multicolumn{4}{c}{Train 1}
& \multicolumn{4}{c}{Train 2} \\
\cmidrule(lr){2-5}
\cmidrule(lr){6-9}
\cmidrule(lr){10-13}
& MD & MR & DA & RA
& MD & MR & DA & RA
& MD & MR & DA & RA \\
\midrule
G1 & 4.488 & 97.253 & 0.299 & 0.406
   & 4.868 & 100.679 & 0.561 & 0.659
   & 5.055 & 102.783 & 0.669 & 0.771 \\
G2 & 4.527 & 97.130 & 0.311 & 0.410
   & 4.762 & 99.086  & 0.479 & 0.546
   & 4.946 & 101.444 & 0.594 & 0.645 \\
G3 & 4.434 & 96.820 & 0.305 & 0.407
   & 4.882 & 101.242 & 0.643 & 0.726
   & 5.036 & 103.645 & 0.748 & 0.817 \\
G4 & 4.554 & 96.707 & 0.330 & 0.435
   & 4.945 & 101.412 & \textbf{0.678} & \textbf{0.741}
   & 5.101 & 104.500 & \textbf{0.800} & \textbf{0.861} \\
G5 & 4.606 & 96.810 & 0.319 & 0.429
   & 4.894 & 100.985 & 0.601 & 0.681
   & 5.036 & 102.081 & 0.677 & 0.764 \\
\bottomrule
\end{tabular}%
}

\parbox{\textwidth}{%
\footnotesize
\textit{Note.} MD = Mean Depth; MR = Mean Rate;
DA = Depth Accuracy; RA = Rate Accuracy.
Except for the pretest, the highest DA and RA values
in each phase are shown in bold.
}
\end{table*}

\begin{table*}[htbp]
\ContinuedFloat
\centering
\caption{(Continued) Performance of Each Group Across the Third Training Round, Posttest, and Delayed Posttest}
\Description{Continuation of the performance table, reporting Train 3, immediate Posttest, and Delayed Posttest results for groups G1 to G5. For each phase, the table lists mean depth, mean rate, depth accuracy, and rate accuracy. G4 has the highest depth and rate accuracy in Train 3 and the immediate Posttest, while G5 has the highest depth and rate accuracy in the Delayed Posttest.}

\resizebox{\textwidth}{!}{%
\begin{tabular}{lcccccccccccc}
\toprule
\multirow{2}{*}{Group}
& \multicolumn{4}{c}{Train 3}
& \multicolumn{4}{c}{Posttest}
& \multicolumn{4}{c}{Delayed Posttest} \\
\cmidrule(lr){2-5}
\cmidrule(lr){6-9}
\cmidrule(lr){10-13}
& MD & MR & DA & RA
& MD & MR & DA & RA
& MD & MR & DA & RA \\
\midrule
G1 & 5.143 & 104.376 & 0.748 & 0.831
   & 4.949 & 100.413 & 0.595 & 0.668
   & 4.801 & 97.750  & 0.497 & 0.573 \\
G2 & 5.064 & 102.368 & 0.686 & 0.731
   & 5.000 & 101.071 & 0.617 & 0.678
   & 4.907 & 98.810  & 0.542 & 0.603 \\
G3 & 5.125 & 103.865 & 0.795 & 0.861
   & 5.025 & 102.025 & 0.713 & 0.773
   & 4.920 & 99.541  & 0.631 & 0.702 \\
G4 & 5.257 & 105.639 & \textbf{0.884} & \textbf{0.916}
   & 5.122 & 103.252 & \textbf{0.818} & \textbf{0.851}
   & 5.036 & 101.135 & 0.730 & 0.779 \\
G5 & 5.112 & 102.629 & 0.750 & 0.806
   & 5.152 & 102.356 & 0.785 & 0.826
   & 5.125 & 100.829 & \textbf{0.748} & \textbf{0.797} \\
\bottomrule
\end{tabular}%
}

\parbox{\textwidth}{%
\footnotesize
\textit{Note.} MD = Mean Depth; MR = Mean Rate;
DA = Depth Accuracy; RA = Rate Accuracy.
Except for the pretest, the highest DA and RA values
in each phase are shown in bold.
}
\end{table*}

\newpage
\begin{table*}[t]
\centering
\caption{Study 1 ANCOVA results for the accuracy-based measures. Groups 1--4 were analyzed in a 2 $\times$ 2 design with Visual condition (ex-situ vs.\ in-situ) and Kinesthetic support (no vs.\ yes) as fixed factors. For the immediate post-test, Post-test was the dependent variable, and Pre-test was entered as the covariate. For delayed retention, Delayed Post-test was the dependent variable, and Post-test was entered as the covariate.}
\Description{Table reporting Study 1 ANCOVA results for groups G1 to G4 in a 2 by 2 design with Visual condition (ex-situ versus in-situ) and Kinesthetic support (absent versus present) as fixed factors. The table is divided into two sections: immediate post-test, where Post-test is the dependent variable and Pre-test is the covariate, and delayed retention, where Delayed Post-test is the dependent variable and Post-test is the covariate. For each section, results are shown for depth accuracy and rate accuracy, including the main effects of Visual and Kinesthetic support, their interaction, and the covariate, with F statistics, p values, and partial eta squared. At the immediate post-test, both Visual condition and Kinesthetic support significantly affect depth accuracy and rate accuracy, and the Visual by Kinesthetic interaction is also significant for both outcomes. The largest effect is Kinesthetic support, especially for depth accuracy, with a very large effect size. The covariates are also significant for both measures. For delayed retention, the main effects of Visual condition and Kinesthetic support are not significant for either depth accuracy or rate accuracy. However, the Visual by Kinesthetic interaction remains significant for both outcomes, with small-to-moderate effect sizes. In this section as well, the covariates are highly significant and account for substantial variance, particularly for rate accuracy. Overall, the table indicates strong immediate effects of both intervention factors and weaker delayed effects that are primarily reflected in their interaction rather than in either main effect alone.}
\label{tab:study1_ancova_rate}
\begin{tabular}{llcccc}
\toprule
Outcome & Effect & $F(df_1, df_2)$ & $p$ & Partial $\eta^2$ \\
\midrule
\multicolumn{5}{l}{\textit{Immediate post-test (DV = Post-test, covariate = Pre-test)}} \\
\midrule
Depth accuracy & Visual & $32.75(1, 43)$ & $< .001$ & .432 \\
Depth accuracy & Kinesthetic & $511.68(1, 43)$ & $< .001$ & .922 \\
Depth accuracy & Visual $\times$ Kinesthetic & $64.25(1, 43)$ & $< .001$ & .599 \\
Depth accuracy & Covariate (Pre-test depth accuracy) & $50.00(1, 43)$ & $< .001$ & .538 \\
\addlinespace
Rate accuracy & Visual & $10.69(1, 43)$ & .002 & .199 \\
Rate accuracy & Kinesthetic & $233.98(1, 43)$ & $< .001$ & .845 \\
Rate accuracy & Visual $\times$ Kinesthetic & $19.16(1, 43)$ & $< .001$ & .308 \\
Rate accuracy & Covariate (Pre-test rate accuracy) & $20.14(1, 43)$ & $< .001$ & .319 \\
\midrule
\multicolumn{5}{l}{\textit{Delayed retention (DV = Delayed Post-test, covariate = Post-test)}} \\
\midrule
Depth accuracy & Visual & $3.39(1, 43)$ & .073 & .073 \\
Depth accuracy & Kinesthetic & $0.22(1, 43)$ & .639 & .005 \\
Depth accuracy & Visual $\times$ Kinesthetic & $4.59(1, 43)$ & .038 & .096 \\
Depth accuracy & Covariate (Post-test depth accuracy) & $65.99(1, 43)$ & $< .001$ & .605 \\
\addlinespace
Rate accuracy & Visual & $3.00(1, 43)$ & .091 & .065 \\
Rate accuracy & Kinesthetic & $2.65(1, 43)$ & .111 & .058 \\
Rate accuracy & Visual $\times$ Kinesthetic & $5.20(1, 43)$ & .028 & .108 \\
Rate accuracy & Covariate (Post-test rate accuracy) & $142.94(1, 43)$ & $< .001$ & .769 \\
\bottomrule
\end{tabular}
\end{table*}

\begin{table*}[t]
\centering
\caption{Study 2 ANCOVA results for the accuracy-based measures. Group 4 (in-situ + kinesthetic) was compared with Group 5 (in-situ + kinesthetic + fading). For the immediate post-test, Post-test was the dependent variable, and Pre-test was entered as the covariate. For delayed retention, Delayed Post-test was the dependent variable, and Post-test was entered as the covariate.}
\Description{Table reporting Study 2 ANCOVA results comparing Group 4, which received in-situ visual feedback and kinesthetic guidance, with Group 5, which received the same intervention plus adaptive fading. The table is divided into immediate post-test and delayed retention analyses. For the immediate post-test, Post-test is the dependent variable and Pre-test is the covariate. For delayed retention, Delayed Post-test is the dependent variable and Post-test is the covariate. For each analysis, the table reports results for depth accuracy and rate accuracy, including the main effect of condition and the corresponding covariate, with F statistics, p values, and partial eta squared. At the immediate post-test, condition significantly affects both depth accuracy and rate accuracy, with Group 4 outperforming Group 5. The effect is especially large for depth accuracy. The Pre-test covariates are also significant for both outcomes. At delayed retention, condition again significantly affects both depth accuracy and rate accuracy, but in the opposite direction, indicating better retained performance for Group 5 than Group 4. The delayed effect is particularly strong for rate accuracy. The Post-test covariates are also highly significant, especially for delayed rate accuracy. Overall, the table shows that adding adaptive fading reduced immediate post-test performance relative to Group 4 but improved delayed retention for both depth and rate accuracy.}
\label{tab:study2_ancova_rate}
\begin{tabular}{llcccc}
\toprule
Outcome & Effect & $F(df_1, df_2)$ & $p$ & Partial $\eta^2$ \\
\midrule
\multicolumn{5}{l}{\textit{Immediate post-test (DV = Post-test, covariate = Pre-test)}} \\
\midrule
Depth accuracy & Condition (G4 vs.\ G5) & $46.60(1, 21)$ & $< .001$ & .689 \\
Depth accuracy & Covariate (Pre-test depth accuracy) & $13.52(1, 21)$ & .001 & .392 \\
\addlinespace
Rate accuracy & Condition (G4 vs.\ G5) & $10.13(1, 21)$ & .004 & .325 \\
Rate accuracy & Covariate (Pre-test rate accuracy) & $11.40(1, 21)$ & .003 & .352 \\
\midrule
\multicolumn{5}{l}{\textit{Delayed retention (DV = Delayed Post-test, covariate = Post-test)}} \\
\midrule
Depth accuracy & Condition (G4 vs.\ G5) & $21.39(1, 21)$ & $< .001$ & .505 \\
Depth accuracy & Covariate (Post-test depth accuracy) & $18.15(1, 21)$ & $< .001$ & .464 \\
\addlinespace
Rate accuracy & Condition (G4 vs.\ G5) & $40.08(1, 21)$ & $< .001$ & .656 \\
Rate accuracy & Covariate (Post-test rate accuracy) & $119.83(1, 21)$ & $< .001$ & .851 \\
\bottomrule
\end{tabular}
\end{table*}

\begin{figure}[H]
    \centering
    \includegraphics[width=1\linewidth]{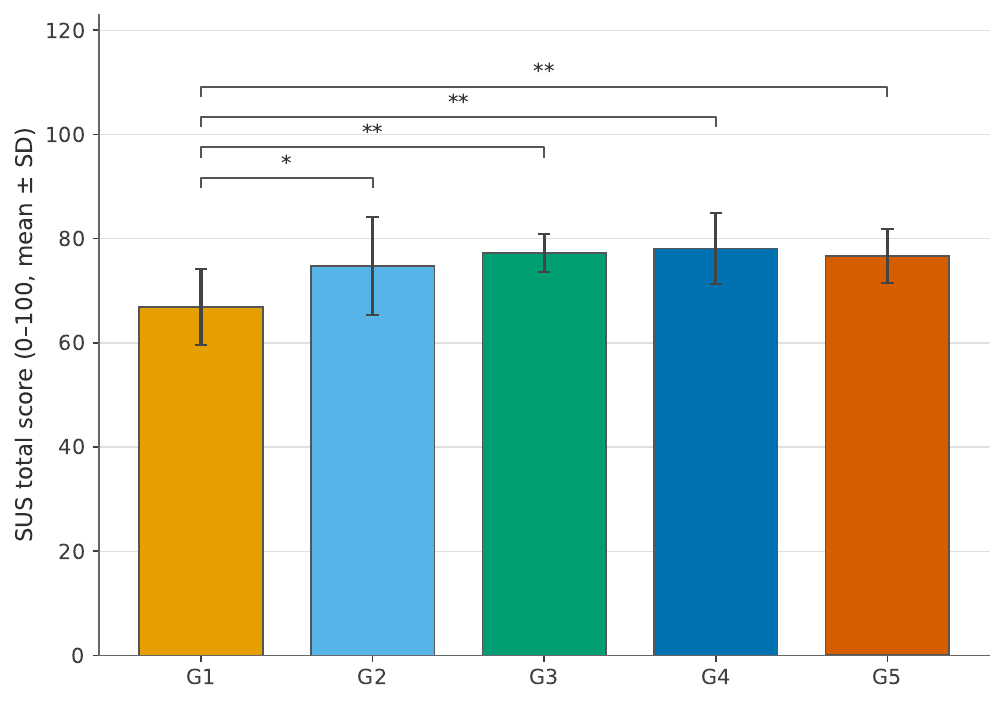}
    \caption{SUS Questionnaire Result.}
    \Description{Bar chart showing mean SUS total scores with standard deviation error bars for five groups, G1 through G5. Scores increase from G1 to G4, with G1 lowest at about 67 and G4 highest at about 78; G5 is slightly below G4 at about 77. Significance brackets indicate that G1 scored significantly lower than G2, G3, G4, and G5, while the differences among G2 to G5 are not marked as significant. Overall, the intervention conditions yielded higher perceived usability than the baseline condition, with the highest usability observed in G4.}
    \label{fig:sus}
\end{figure}

\begin{figure}[H]
    \centering
    \includegraphics[width=1\linewidth]{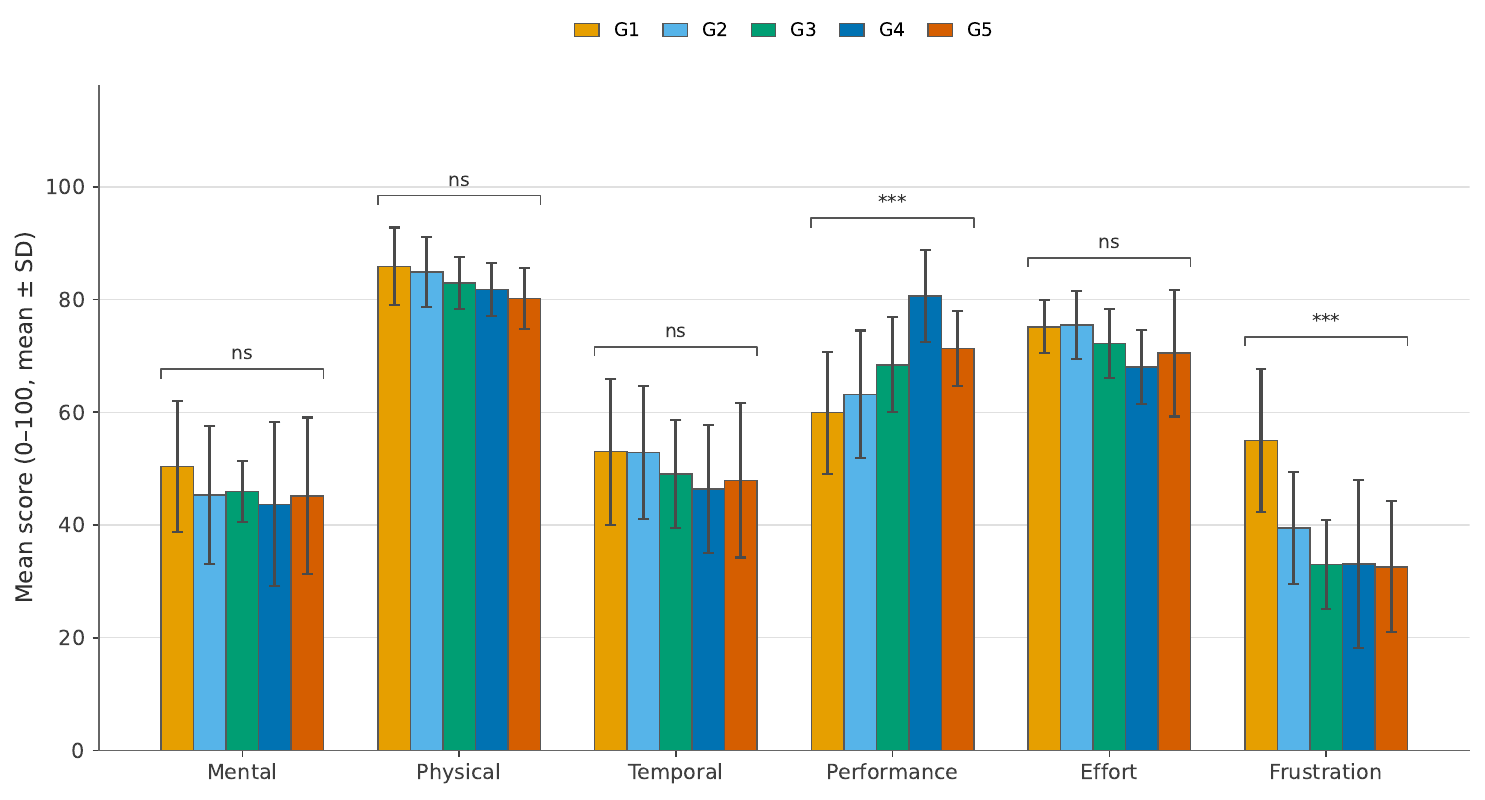}
    \caption{NASA-TLX Questionnaire Result.}
    \Description{Grouped bar chart showing NASA-TLX scores for five groups, G1 through G5, across six workload dimensions: Mental, Physical, Temporal, Performance, Effort, and Frustration. The y-axis reports mean scores from 0 to 100 with standard deviation error bars. Mental, Physical, Temporal, and Effort scores are broadly similar across groups, and each of these categories is marked as not significant. Performance scores differ significantly, with G4 showing the highest mean and G5 also higher than the earlier groups. Frustration scores also differ significantly, with G1 highest and the remaining groups substantially lower, especially G3 to G5. Overall, the chart indicates that the intervention conditions mainly affected perceived performance and frustration rather than the other workload dimensions.}
    \label{fig:nasatlx}
\end{figure}

\end{document}